\documentclass[conference]{IEEEtran}
\usepackage{cite}
\usepackage{graphicx}
\usepackage{amsmath}
\usepackage{times}
\usepackage{latexsym}
\usepackage{graphicx}
\usepackage{bm}
\usepackage{amssymb}
\usepackage[center]{caption2}
\usepackage{stfloats}
\usepackage{cases}
\usepackage{array}
\usepackage{setspace}
\usepackage{fancyhdr}
\usepackage{amscd}
\usepackage{enumerate}
\usepackage{multirow}
\usepackage{float}
\usepackage{color}
\usepackage{longtable}
\usepackage{algorithm}
\usepackage[caption=false]{subfig}

\usepackage{enumitem}
\usepackage[noend]{algorithmic}
\graphicspath{ {./img/png/} }
\allowdisplaybreaks

\newcommand{\Pidle}{P_{\mathrm{idle}}^a}
\IEEEoverridecommandlockouts
\begin{document}
\thispagestyle{empty}
{\title{\color{black}Virtualization of Multi-Cell 802.11 Networks: Association and Airtime Control}}
\author{
    \IEEEauthorblockN{ Mahsa Derakhshani\IEEEauthorrefmark{2}, Xiaowei Wang\IEEEauthorrefmark{1}, Tho Le-Ngoc\IEEEauthorrefmark{1}, Alberto Leon-Garcia\IEEEauthorrefmark{2}}
     \IEEEauthorblockA{\IEEEauthorrefmark{2}Department of Electrical \& Computer Engineering, University of Toronto, Toronto, ON, Canada}
    \IEEEauthorblockA{\IEEEauthorrefmark{1}Department of Electrical \& Computer Engineering, McGill University, Montreal, QC, Canada}
    \IEEEauthorblockA{Email: mahsa.derakhshani@utoronto.ca; xiaowei.wang@mail.mcgill.ca; \\tho.le-ngoc@mcgill.ca; alberto.leongarcia@utoronto.ca}
}
\providecommand{\keywords}[1]{\textbf{\textit{Index terms---}} #1}
\maketitle
\begin{abstract}

This paper investigates the virtualization and optimization of a multi-cell WLAN.  We consider the station (STA)-access point (AP) association and airtime control for virtualized 802.11 networks to provide service customization and fairness across multiple internet service providers (ISPs) sharing the common physical infrastructure and network capacity. More specifically, an optimization problem is formulated on the STAs' transmission probabilities to maximize the overall network throughput, while providing airtime usage guarantees for the ISPs. Subsequently, an algorithm to reach the optimal solution is developed by applying monomial approximation and geometric programming iteratively. Based on the proposed three-dimensional Markov-chain model of the enhanced distributed channel access (EDCA) protocol, the detailed implementation of the optimal transmission probability is also discussed. The accuracy of the proposed Markov-chain model and the performance of the developed association and airtime control scheme are evaluated through numerical results.
\end{abstract}
\section{Introduction}\label{Section-Introduction}
\subsection{\color{black}Background and Motivation}
Wireless virtualization has recently emerged as an architectural choice for the wireless networks, in which different service providers can share physical infrastructure and wireless resources. Allowing resources to be shared, vitualization can facilitate a flexible and lower-complexity solution to support customized services with finer control over quality-of-service (QoS) features. To enable service customization, the key issue would be to provide isolation among virtual networks running by different service providers. Such isolation can be achieved through contention-free resource allocation techniques based on TDMA and/or FDMA, by isolating resources across service providers. But, in systems using contention-based access protocols, providing isolation among virtual networks is more challenging. 

For instance, in virtualized 802.11 WLANs, transmissions of different virtual WLANs (V-WLANs) are closely coupled, although administrative virtualization (i.e., one physical AP advertises multiple service set identifiers (SSIDs)) can already differentiate groups of flows. With a carrier sense multiple access (CSMA)-based MAC, unavoidable collisions act to couple the transmissions of different V-WLANs. Moreover, since the network capacity is shared yet constrained, the increase of traffic in one V-WLAN may reduce the available network capacity to another \cite{checcofair}. Thus, an efficient resource allocation among V-WLANs is essential to manage the MAC-layer couplings.


In a 802.11 WLAN with densely deployed APs, before a STA can access the network, it needs to make a decision about which AP to associate with. In virtualized 802.11 networks, such STA-AP association control could create an opportunity to provide fairness guarantees among different ISPs. {\color{black}In addition to STA-AP association, the airtime among STAs associated to the same AP can optimally be adjusted. Controlling airtime usage of the STAs provides the opportunity to optimize the WLAN performance (e.g., improving WLAN throughput by exploiting multi-user diversity) as well as another degree of freedom to guarantee fairness among the ISPs.
}


{\color{black}The main focus of this work is on the virtualization and optimization of a multi-cell WLAN. More specifically, }STA-AP association and airtime control are jointly explored to provide fairness and throughput guarantees for different V-WLANs. Taking into account STA transmission rates and ISP airtime reservations, an optimization problem is formulated to adjust the transmission probability of each STA at each AP. The objective is to maximize the overall network throughput, while keeping a total airtime guarantee for each ISP. To solve the formulated problem which is non-convex and thus computationally intractable, an iterative algorithm is developed through successive geometric programming. This algorithm can achieve an optimal solution with an affordable complexity.

Furthermore, to implement the optimal transmission probabilities, we model the EDCA protocol with a three-dimensional Markov chain and establish the relationship between the transmission probability of a STA and the EDCA parameters. Based on this relationship, we developed a control algorithm to approach the optimal transmission probability by {\color{black} jointly} manipulating the EDCA parameters such as contention window size (CW) and arbitration inter-frame space (AIFS). Finally, through numerical results, we verify the performance enhancement provided by the developed STA-AP association and airtime control approach in terms of throughput and fairness.

\subsection {\color{black}Related Works}
\subsubsection{STA-AP Association and Airtime Control}
In most current vendor implementations, 802.11 STAs choose the AP with the highest received signal-to-noise ratio (SNR) to connect with. Since the STA density is often uneven in the network \cite{2003_Balazinska_Mobility,2004_Schwab_Campus_WLAN}, the Max-SNR approach can lead to an unbalanced distribution of STAs among APs, causing unfairness. In order to balance the load of APs, several STA-AP association algorithms have been presented in the literature, mostly by maximizing the minimum throughput of all STAs  \cite{2008_Huazhi_DistributedFairness,2008_Ercetin_Game_1,2010_Wenchao_Game_2,2012_Pradeepa_Web_Delay}. Nevertheless, in a basic service set (BSS) including an AP and its associated multi-rate STAs, it is shown that the throughput is limited by the STA with the lowest data rate. This phenomenon is also \textcolor{black} to as the \textit{performance anomaly} problem \cite{2003-Heusse-PerformanceAnomaly}. Thus, comparing with the Max-SNR approach, these load-balancing approaches improve the max-min fairness among STAs at the cost of decreasing the aggregate throughput.
 
To address the \textit{performance anomaly} and balance the trade-off between aggregate throughput and fairness, proportional fair throughput allocation has widely been considered in multi-rate 802.11 WLANs \cite{2011-Checco-ProportionalFairness,2012_Xiaomin_Coding,2008_Soung_PropotionalFairness_Part1,2008_Soung_PropotionalFairness_Part2,2010_Hui_Global}. In \cite{2011-Checco-ProportionalFairness}, proportional fairness is studied in a single BSS. It is shown that \textit{propotional fairness} leads to an \textit{airtime-fairness}, where equal airtime usage is provided to all STAs. Moreover, in a multi-AP WLAN, \cite{2008_Soung_PropotionalFairness_Part1,2008_Soung_PropotionalFairness_Part2} study STA-AP association problem with an objective to maximize the proportional fairness. More precisely, association control is implemented in a form of airtime allocation, where the transmission time of STAs at different APs are jointly optimized \cite{2008_Soung_PropotionalFairness_Part1,2008_Soung_PropotionalFairness_Part2}. 

Precise controlling of the STA airtime usage in a 802.11 contention-based WLAN is very difficult due to the distributed and random nature of their CSMA-based MAC protocol. Nevertheless, the airtime control algorithms have been proposed in the literature by manipulating the MAC parameters \cite{Joshi2008AirTimeFariness802.11,Chun2006AirTimeFariness802.11e,2010_Laddomada_Traffic,2011_Pochiang_Delay_Sensitive}. For instance, \cite{Chun2006AirTimeFariness802.11e} discusses how to dynamically modify the CW based on the transmission data rate and estimated packet error rate (PER) in a WLAN aiming to guarantee airtime fairness. Similarly, in \cite{2010_Laddomada_Traffic}, CW is adjusted according to the traffic load of each STA to improve throughput and fairness in the WLAN. Furthermore, \cite{2011_Pochiang_Delay_Sensitive} proposes a machine-learning approach to adjust AIFS and CW aiming to achieve airtime fairness and reduce frame delay. Such heuristic works provide pretty good understanding of the influence of specific MAC parameters such as CW and AIFS on the transmission probability and thus airtime usage of the STAs. However, the discussions are limited to adjusting the parameters separately and thus can only find a suboptimal solution in a subset of the feasibility region. 

In a virtualized WLAN serving multiple ISPs, STA-AP association and airtime control become more challenging. {\color{black}The reason is that fairness guarantees and service customization are required for ISPs}, while there are unavoidable couplings among the STA transmissions of different ISPs in the network. There are a few works addressing only airtime control in the literature. Considering a virtualized {\color{black}single-cell} WLAN, in \cite{2012_Nakauchi_Airtime}, a heuristic airtime control algorithm is proposed to achieve the target airtime usage for each ISP by controlling the minimum CW of each STA. Similarly, \cite{banchs2012providing} addresses optimizing CW using control theory. But, as the discussion is limited to controlling minimum CW, the optimality of the result might be sacrificed. In a {\color{black}multi-cell} WLAN, \cite{checcofair} presents an analysis on the feasibility region of ISP airtimes and characterizes the ISP airtimes at the rate region boundary. Furthermore, a distributed algorithm is developed to allocate airtime slices among ISPs and flow rates within each slice in a max-min fair manner. Since max-min fairness is used as an objective for rate allocation among the flows in each ISP, the optimality of the achieved total throughput may not be guaranteed in a multi-rate WLAN. In addition, the association control is not discussed in \cite{checcofair}.

\subsubsection{IEEE EDCA Modeling and Optimization }
{\color{black}
To implement the optimal STA-AP association and airtime allocation, it is desirable to control the transmission probability of each STA. However, in a CSMA/CA-based WLAN, the only controllable parameters are the MAC layer parameters. Thus, it is essential to mathematically model the effects of such parameters on the transmission probabilities of STAs.

The EDCA modeling has received considerable attention in the literature. There are several studies, such as \cite{2008_Ramaiyan_Fixed,2010_Tinnirello_Rethinking}, that only focus on numerically solving the stationary state (transmission probability of each STA) given the network configuration. Another group of works, e.g., \cite{2004_Robinson_Saturation,2006_Banchs_Throughput,2014_Parker_Increasing,dong2015boosting,2009_Jae_Control,2014_Gao_EDCA,2007_Bo_Achieving,2007_Ge_Analytic}, provide models for the EDCA protocol with the help of a Markov chain, inspired by the classical work of Bianchi \cite{Bianchi2000}. Using such Markov chain models, it is possible to optimize the network by tuning the EDCA parameters. However, due to the complexity of the proposed EDCA models, it is hard to establish an explicit relationship between the transmission probabilities and the controllable EDCA parameters. Thus, these works mostly provide a numerical (e.g., \cite{2004_Robinson_Saturation,2006_Banchs_Throughput,2014_Parker_Increasing,dong2015boosting}) or approximate analytical solution in terms of MAC parameters (e.g., \cite{2009_Jae_Control,2014_Gao_EDCA,2007_Bo_Achieving,2007_Ge_Analytic}). Moreover, none of these works has jointly controlled the EDCA parameters. 


Our proposed Markov chain model is based on \cite{Kong_Makov_802.11e}, with which an explicit relationship between the transmission probability and EDCA MAC parameters can be established. One limitation of the EDCA model proposed in \cite{Kong_Makov_802.11e} is the accuracy of AIFS differentiation. Unlike in DCF, the time duration each STA has to wait before its backoff process (i.e. AIFS) is different in EDCA. This can lead to the result that the number of contending STAs is not time-homogeneous \cite{2010_Tinnirello_Rethinking}. Therefore, the collision probability that each STA faces is also not time-homogeneous. But, as pointed out in \cite{2014_Gao_EDCA}, time-homogeneity assumption greatly simplifies the modeling complexity, and thus, the steady state performance can be characterized as explicit functions of backoff parameters. Furthermore, the accuracy of such model can be effectively improved by setting a relatively large initial CW.
   
\subsection{\color{black}Structure}
The rest of this paper is organized as follows. Section \ref{Section-System-model} presents an overview of the system configuration and modeling. 
In Section \ref{Section-Optimization-Problem}, we first analyze the feasibility region of the transmission probabilities based on the proposed Markov chain model for IEEE EDCA. Then, we formulate the transmission probability optimization problem, which maximizes the system throughput and guarantees the fairness among the ISPs. In Section \ref{Section:Implementation-and-Numerical-Result}, the implementation details of MAC parameter control are discussed in order to achieve the optimal transmission probability. Illustrative results are provided in Section \ref{Section:Illustrative-Results} to evaluate the performance of the developed algorithms. Section \ref{Section:Conclusion} provides some concluding remarks.  
}
\section{System Configuration and Modeling}\label{Section-System-model}
\begin{figure}[!t]
\centering
\includegraphics[width=0.95\linewidth]{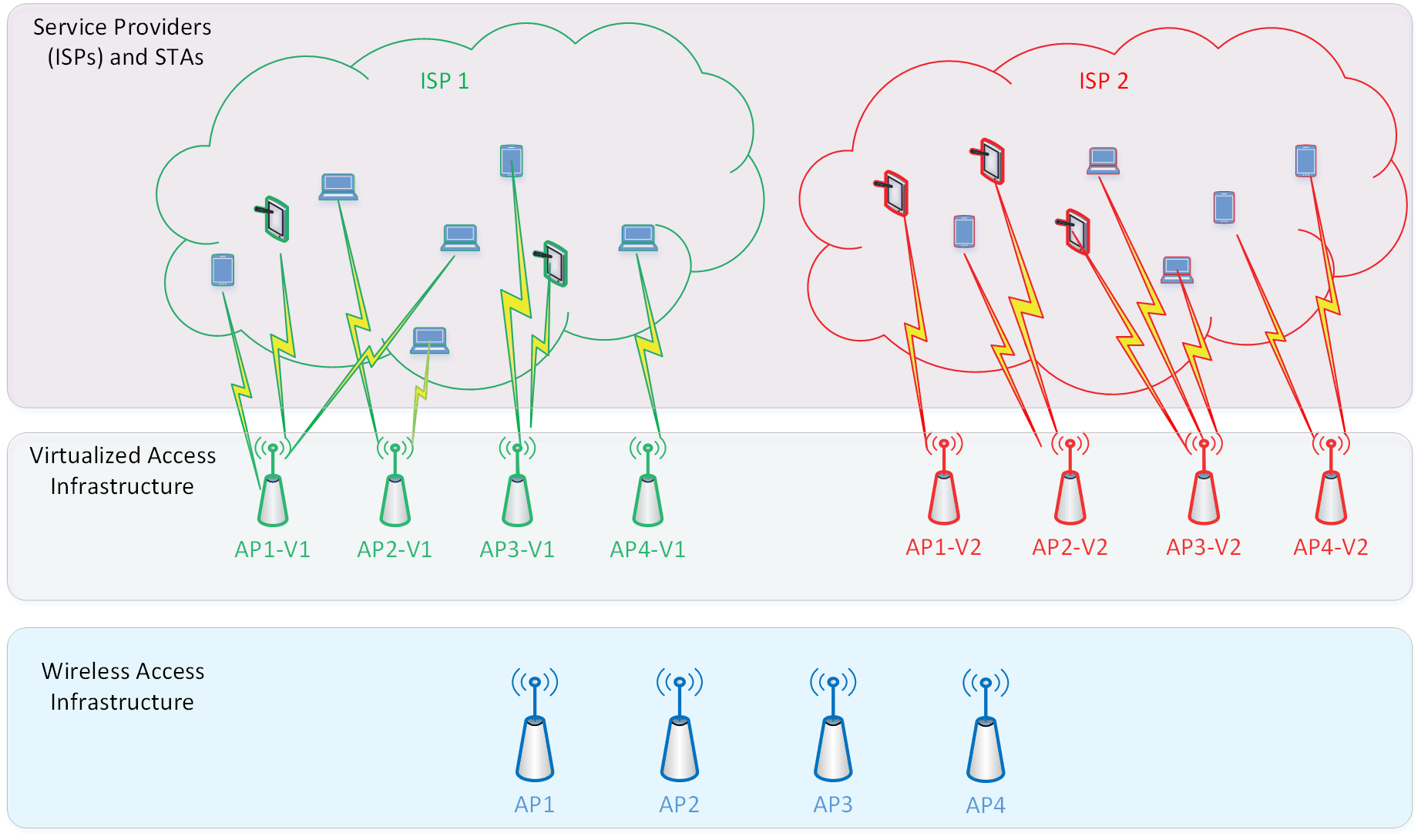}
\caption{Layered system model}
\label{Fig:Layered-System-Model}
\vspace{-3mm}
\end{figure}
\begin{figure}[!t]
\centering
\includegraphics[width=0.95\linewidth]{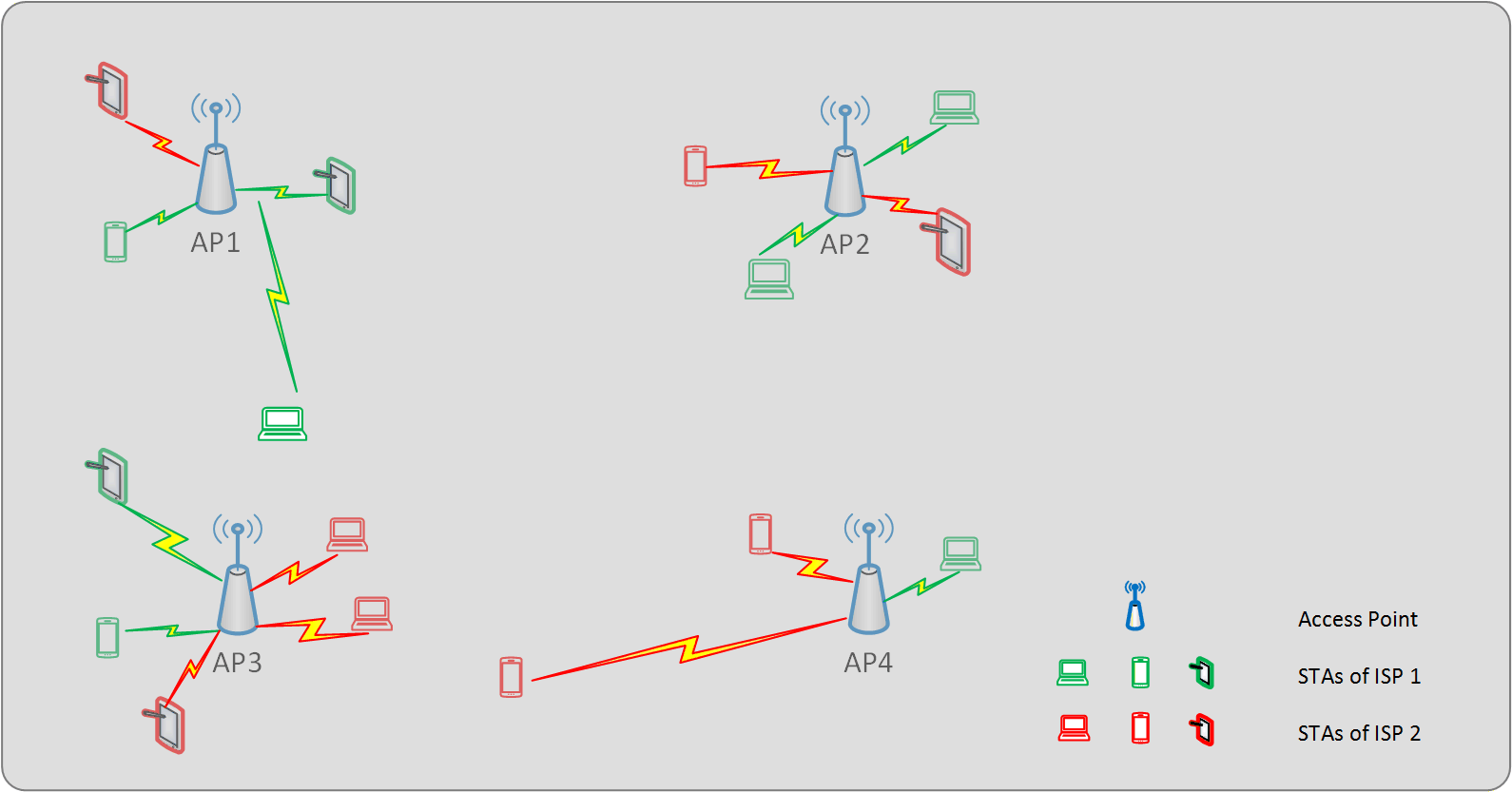}
\caption{System model}
\label{Fig:System_Model}
\vspace{-5mm}
\end{figure}
We consider an IEEE 802.11-based WLAN that consists of a large number of APs. APs operate on non-overlapping frequency channels. Let $\mathcal{A}$ be the set of APs and $N_a=|\mathcal{A}|$ be the total number of APs. Each AP has a limited coverage area and all STAs are randomly distributed in the field. The network carries traffic belonging to a number of different ISPs (also referred to as V-WLANs). Let $\mathcal{K}$ be the set of ISPs using the network. Furthermore, let $\mathcal{S}_k$ be the set of STAs of ISP $k \in \mathcal{K}$ and $N_k=|\mathcal{S}_k|$ be the number of STAs belong to ISP $k$. Furthermore, let $\mathcal{S}$ be the set of all STAs and $N_s=\sum_{k \in \mathcal{K}} N_k$ be the total number of STAs in the network. The network is administratively virtualized, i.e., each AP will broadcast multiple different SSIDs, one for each ISP. Figure \ref{Fig:Layered-System-Model} illustrates an example of the network architecture with four physical APs and two ISPs.
\subsection{Enhanced Distributed Channel Access (EDCA)}
To access the channel, the STAs are assumed to follow the EDCA MAC protocol of 802.11e standard. EDCA protocol is designed to enhance the basic MAC mechanism, i.e., distributed coordination function (DCF), aiming to support service differentiation. Similar to DCF, EDCA is also a contention-based access scheme, based on CSMA/CA using binary exponential backoff rules to manage retransmission of collided packets. Here, we briefly review the EDCA operation, as standardized by 802.11 protocol \cite{xiao2004ieee,Derakhshani2014,Bianchi2000}. 

EDCA requires a STA to monitor the channel before any transmission. If the channel is sensed idle for a time interval equal to an arbitration inter-frame space (AIFS), the STA transmits. Otherwise, if the STA senses a transmission either immediately or during the AIFS, it continues monitoring the channel. When the channel is measured idle for an AIFS, the STA backoffs for a random period of time. EDCA uses a discrete-time backoff mechanism, i.e., the time following an AIFS is slotted. The backoff time is selected according to a uniform distribution in the interval $[0,W_i]$ where $W_i$ represents the contention window of STA $i \in \mathcal{S}$\cite{xiao2004ieee,Derakhshani2014,Bianchi2000}. 

STA $i$ starts a packet transmission at backoff stage 0, where $W_i$ is set equal to the minimum contention window size $W_{\min,i}$. Then, after each unsuccessful transmission, STA $i$ moves to the next backoff stage and doubles $W_i$. When the maximum backoff stage $m_i$ is reached, $W_i$ is no longer increased and stays at $2^{m_i}W_{\textrm{min},i}$. If the STA experiences a collision at the backoff stage $m_i$, it will retry transmission for at most $h_i$ times, where $h_i$ is the retransmission limit at the maximum backoff stage. If the packet is still not successfully transmitted after $h_i$ retransmissions, it will be discarded \cite{xiao2004ieee,Derakhshani2014,Bianchi2000}.

\begin{figure}[t]
\centering
\includegraphics[width=1\linewidth]{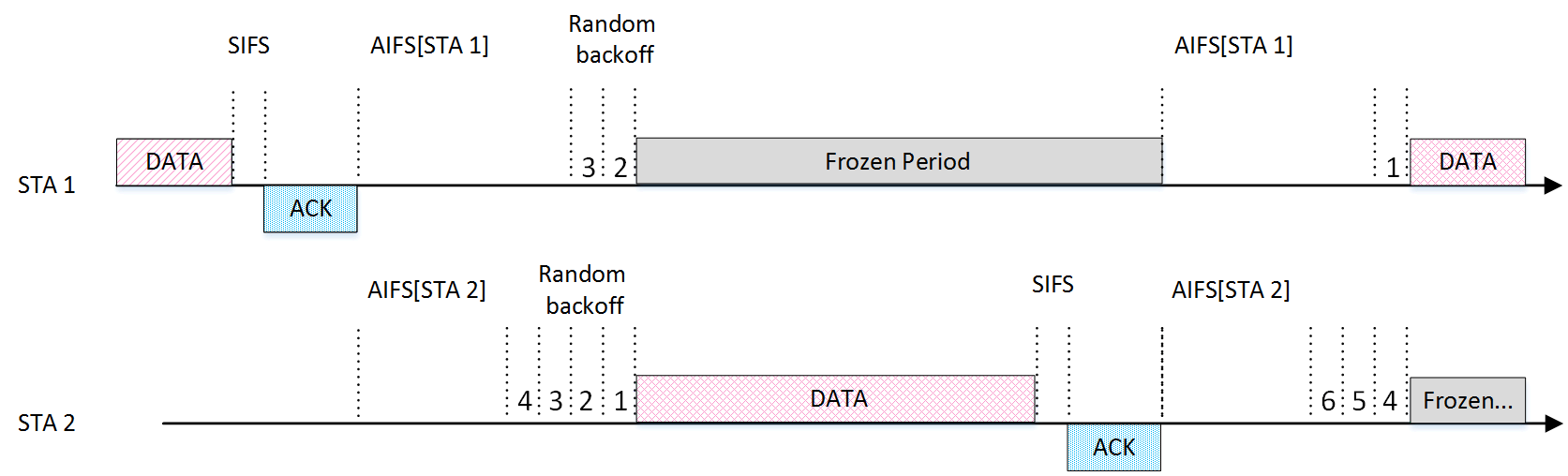}
\caption{IEEE 802.11e EDCA channel access procedure}
\label{Fig:802.11e}
\vspace{-2mm}
\end{figure}

The backoff time counter is decremented and a STA transmits when the backoff counter reaches zero. If the data frame is successfully received, the AP waits for a period of time called short inter-frame space (SIFS) and then sends an acknowledgment (ACK). Once the channel is sensed busy while counting down, the STA will freeze its backoff counter and continue decrementing when channel finds idle again. Figure \ref{Fig:802.11e} illustrates an example of the channel-access procedure of two STAs using EDCA \cite{xiao2004ieee,Derakhshani2014,Bianchi2000}.

EDCA allows to have variable waiting times and contention parameters (i.e., AIFS and $W_{\min}$) to enable service differentiation. Using EDCA, for each transmission opportunity, a STA can transmit multiple back-to-back packets for a fixed period of time. Let $T_{\textrm{TXOP}}$ be the duration of a data frame. Then, the duration of a successful transmission becomes
\begin{align}
T_s=T_{\textrm{TXOP}}+\text{SIFS}+\gamma+\text{ACK}+\gamma+\text{AIFS}
\end{align}
where $\gamma$ denotes the transmission delay. Similarly, the duration of a collision can be calculated as
\begin{align}\label{T_c_Definition}
T_c=T_{\textrm{TXOP}}+\gamma+\text{AIFS}.
\end{align}

{
\color{black}
\noindent
Note that when the colliding STAs use different AIFS values, the AIFS value in (\ref{T_c_Definition}) should take the value of the largest AIFS. But, since the fixed transmission duration $T_{\text{TXOP}}$ (in the order of $ms$) is much larger than AIFS in the order of $\mu s$, the difference between the AIFS can be ignored here.
}
%
\subsection{Association Control via Transmission Probabilities}
In a WLAN with APs densely deployed, STAs need to determine which APs to connect with. We aim to generalize the association control problem by adjusting the transmission probability of each STA at any AP, rather than selecting one AP to associate with. Thus, we define $\tau_i^a$ ($0\leq \tau_i^a\leq 1$) as the probability that STA $i$ attempts to transmit at AP $a$ in a general time-slot. Consequently, the probability that a time-slot is idle in the BSS including AP $a$ is 
\begin{align}
	\Pidle=\prod\nolimits_{i \in \mathcal{S}}(1-\tau_i^a).
\end{align}
In a given BSS, transmitted packets will be received successfully, if exactly one STA transmits on the channel. Thus, the probability of a successful transmission initiated by STA $i$ becomes
\begin{align}
P_{\text{succ},i}^a=\tau_i^a \prod\nolimits _{i' \in \mathcal{S}, i'\neq i}(1-\tau_{i'}^a).
\end{align}
{\noindent
\color{black}
Since $\text{ACK}$ and $\text{SIFS}$ are relatively small (in the order of $\mu\text{s}$) compared with $T_{\textrm{TXOP}}$ (in the order of $\text{ms}$), we approximate $T_s$ and $T_c$ to be of the same size and denote them by $T$. Consequently, the expected length of a general time-slot becomes
}
\begin{align}\label{Eq:ETg}
\mathbb{E}\{T_g\}=\delta\Pidle+(1-\Pidle)T
\end{align}
where $\delta$ is the duration of an idle time-slot. Furthermore, the expected information (in bits) transmitted by STA $i$ to AP $a$ in a general time-slot can be derived as
\begin{align}\label{Eq:EIg}
	\mathbb{E}\{I_g\}=P_{\text{succ},i}^a r_i^a T_{\textrm{TXOP}}
\end{align}
where $r_i^a$ represents the transmission data rate of the link between STA $i$ and AP $a$. As defined in \cite{Bianchi2000}, based on the (\ref{Eq:ETg}) and (\ref{Eq:EIg}), the throughput of STA $i$ at AP $a$ becomes 
\begin{align}
T_i^a=\frac{\mathbb{E}\{I_g\}}{\mathbb{E}\{T_g\}}=\frac{P_{\text{succ},i}^a r_i^a T_{\mathrm{TXOP}}}{\Pidle\delta+(1-\Pidle)T}.
\end{align}
Let define a new variable $x_i^a=\frac{\tau_i^a}{1-\tau_i^a}$ ($x_i^a\geq 0$), which represents the expected number of consecutive transmission attempts by STA $i$ at AP $a$ as \cite{checcofair,2011-Checco-ProportionalFairness}. Consequently, $\Pidle$ and $P_{\text{succ},i}^a$ will be transformed into
\vspace{-1mm}
\begin{align}\label{Eq:Pidle}
\Pidle=\frac{1}{\prod _{i \in \mathcal{S}}(1+x_i^a)},
\end{align}
\begin{align}\label{Eq:Psucc}
P_{\text{succ},i}^a=\frac{x_i^a} {\prod _{i' \in \mathcal{S}}(1+x_{i'}^a)}=x_i^a \Pidle.
\end{align}
Subsequently, from (\ref{Eq:Pidle}) and (\ref{Eq:Psucc}), $T_i^a$ can be represented in terms of $x_i^a$ as
\begin{align}\label{Eq:Throughput}
T_i^a=\frac{x_i^a \Pidle r_i^a T_{\mathrm{TXOP}}}{T-(T-\delta)\Pidle}=\frac{x_i^a r_i^a t}{\prod_{i' \in \mathcal{S}}(1+x_{i'}^a)-t'}
\end{align}
where $t=\frac{T_{\mathrm{TXOP}}}{T}$ and $t'=\frac{T-\delta}{T}$. 

In addition to the throughput of each STA, the fraction of time that each STA spends for transmission could be considered as another performance metric, specifically in order to measure and preserve fairness among different STAs or ISPs. The total access airtime for STA $i$--including both successful transmissions and collisions--becomes
\begin{align}\label{Eq:Airtime}
T_{\mathrm{air},i}^a=\frac{P_{\mathrm{coll},i}^a T+P_{\text{succ},i}^a T}{\Pidle \delta + (1-\Pidle)T}
\end{align}
where $P_{\mathrm{coll},i}^a=\tau_i^a \left[1-\prod_{i' \in \mathcal{S},i'\neq i}(1-\tau_{i'}^a)\right]$ is the probability that STA $i$ suffers from a collision in a general time-slot as defined in \cite{2011-Checco-ProportionalFairness}. Consequently, 
\begin{align}\label{Eq:Airtime2}
T_{\mathrm{air},i}^a=\frac{\tau_i^a}{1-\Pidle t'}=\frac{x_i^a \prod_{i' \in \mathcal{S}, i' \neq i}(1+x_{i'}^a)}{\prod_{i' \in \mathcal{S}}(1+x_{i'}^a)-t'}.
\end{align}
In this work, we aim to maximize the overall network throughput, while guaranteeing a minimum requirement on the aggregate airtime of each ISP. To this end, the transmission probability of STAs ($\tau_i^a$) need to be adaptively optimized by maximizing the aggregate throughput of all STAs at all APs (i.e.,$\sum_{i \in \mathcal{S}, a \in \mathcal{A}}T_i^a$). Furthermore, for each ISP (e.g., ISP $k$), a constraint needs to be set in order to keep the total airtime of all STAs belonging to ISP $k$ larger than a minimum requirement. More specifically, $\sum_{i \in \mathcal{S}_k, a \in \mathcal{A}} T_{\mathrm{air},i}^a \geq \eta_k $ where $\eta_k$ denotes the target share of the airtime for ISP $k$.

{
\color{black}
Accordingly, to formulate such optimization problem, the feasibility region of $\tau_i^a$ (or $x_i^a $) is required. Thus, we study the behavior of a single STA -which is using EDCA- with a three-dimensional Markov chain. With the aid of the proposed Markov model, we can learn how to implement or control $\tau_i^a$ in terms of EDCA parameters. As a result, based on the established relationship between $\tau_i^a$ and EDCA parameters, we would be able to analyze its feasibility region and also design an algorithm to control EDCA parameters to approach the optimal $\tau_i^a$.
}

	\begin{figure*}[t] 
		\begin{minipage}[t]{\linewidth}
			\begin{@twocolumnfalse}
				\begin{align}
					\tau_i&=\sum\limits_{j=0}^{m_i+h_i}b_{j,0,0}=\frac{1-p_i^{m_i+h_i+1}}{1-p_i}\left[L_i\frac{1-q_i}{q_i}+\frac{1+p_iN}{p_i}\frac{1-(1-p_i)^{A_i+1}}{(1-p_i)^{A_i+1}}+\frac{1-p_i^{m_i+h_i+1}}{1-p_i}+
					\right. \label{Tau_In_variables} \\
					&~~~~~~~~~~~~~~~~~~~~~~~~~~~~~~~~~~~~\left. \frac{1+Np_i}{2(1-p_i)^{A_i}}\left(W_{\textrm{min},i}\left[\frac{1-(2p_i)^{m_i+1}}{1-2p_i}+\frac{2^{m_i}p_i^{m_i+1}(1-p_i^{h_i})}{1-p_i}\right]-\frac{1-p_i^{m_i+h_i+1}}{1-p_i}\right)\right]^{-1} \nonumber 
				\end{align}
				\vspace{-5mm}
				\begin{center}
					\line(1,0){515}
				\end{center}
			\end{@twocolumnfalse}
		\end{minipage}
	\end{figure*}
\section{Optimization Problem}\label{Section-Optimization-Problem}
In this section, we present the STA-AP association and airtime control optimization problem based on the system model introduced in Section \ref{Section-System-model}. We first use a Markov chain model to model the EDCA protocol. With this model, we can analyze the feasibility region of $\tau_i^a$ for each STA. Then, the optimization problem is formed and solved by applying monomial approximation and geometric programming.
\subsection{Markov Chain Model for EDCA}\label{SubSection:Markov_Chain}
\begin{figure}[!t]
\centering
\includegraphics[width=1\linewidth]{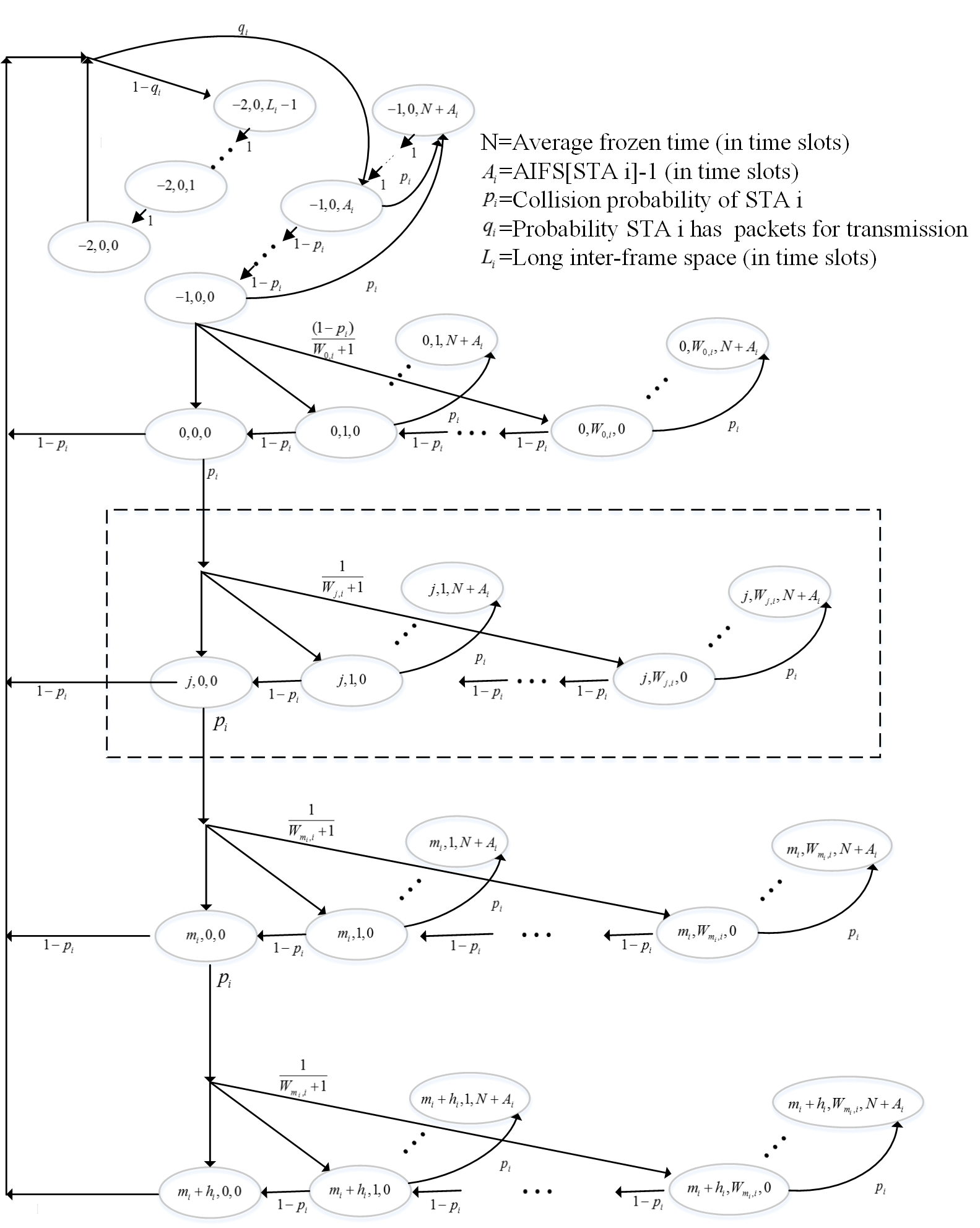}
\caption{Three dimensional Markov chain model for one STA}
\label{Fig:Markov-Chain}
\vspace{-5mm}
\end{figure}
{\color{black}
To study the feasibility region of $\tau_i^a$ and its implementation in the EDCA MAC protocol, we establish a three-dimensional Markov chain. Such model enables us to estimate the transmission probability of the STAs in a WLAN employing the EDCA protocol. Different from the previous works, which discuss a single parameter in the EDCA protocol (e.g., \cite{Chun2006AirTimeFariness802.11e}), we aim at developing a general mathematical model that can show the influences of all the EDCA parameters on the transmission probability.

Figure \ref{Fig:Markov-Chain} shows the three-dimensional Markov chain model to describe the behavior of a single STA, which uses the EDCA MAC protocol. Since STA transmissions are only coupled within a BSS, we consider a single BSS where an AP and its associated STAs reside in. Thus, in this subsection without loss of generality, we remove the index of $a$ (e.g., $\tau_i^a$ is replaced by $\tau_i$) to keep the notation simple. 

Our proposed three-dimensional Markov chain model is an extension to the presented model in \cite{Kong_Makov_802.11e}. To be able to further control the competition among STAs and thus the collision probability in a BSS, we introduce two new MAC parameters (i.e., $q_i$ and $L_i$ for STA $i$) for each STA. Similar to \cite{Mahsa_2013_Adaptive}, after a successful transmission or a packet drop, the STA has to flip a biased coin with successful probability $q_i$ to enter the backoff process. Whenever the STA fails to enter the backoff process, it needs to wait for a period of long inter-frame space (i.e., $L_i$ time-slots) before another try. These two variables are incorporated in the proposed Markov chain.

It should be noted that time in this Markov chain is slotted and the interval between any two adjacent states is a general time-slot. Here, a general time-slot can be an idle time-slot $\delta$, a successful transmission duration ($T_s$) or a collision duration ($T_c$). In this Markov chain, the triple $\{s(t),b(t),v(t)\}$ denotes the state of the STA at time $t$. More specifically, $s(t)$ represents the backoff/retransmission stage of STA, where $-2\leq s(t)\leq m_i+h_i$ ($s(t)=-2,-1$ will be explained later). Furthermore, $b(t)$ denotes the backoff counter, which takes a value between $0$ and the contention window of the current backoff stage. Furthermore, $v(t)$ represents the remaining time of a frozen period or the waiting time to the next trial when the STA already failed to enter the backoff process.

The detailed transition probabilities between states of the Markov chain and the stationary condition of the Markov chain can be found in the Appendix. With such stationary conditions, the transmission probability $\tau$ of each STA is derived as function of the detailed EDCA parameters as in (\ref{Tau_In_variables}), where $p_i$ is the collision probability faced by STA $i$ as defined in (\ref{Eq:Pi}).
}
\subsection{Feasibility Region Analysis}\label{SubSection:Feasibility_Region}
In this subsection, we study the feasibility region of $\tau_i^a$ in order to complete the optimization problem formulation. Based on (\ref{Tau_In_variables}), the adjustable variables to tune $\tau_i^a$ are $m_i^a$, $h_i^a$, $L_i^a$, $q_i^a$, $A_i^a$, and $W_{j,i}^a$ where $0\leq j\leq m_i^a+h_i^a$. Recall that $0 \leq q_i^a \leq 1$, $0 \leq L_i^a$, and $0 \leq W_{j,i}^a$. Furthermore, since $A_i=\text{AIFS}[\text{STA}~i]-1$, and $\text{AIFS}>\text{SIFS}$ in EDCA , we have $A_i \geq \text{SIFS}$. Assuming that $\text{SIFS}$ is equal to one time-slot, we have $ A_i^a\geq 1$ without loss of generality.

\subsubsection{Lower Bound of $\tau_i^a$}
Let $\underline{\tau}^a_i$ be the lower bound of $\tau_i^a$. From (\ref{Tau_In_variables}), it is clear that $\tau_i^a\rightarrow 0$ when $L_i^a\rightarrow\infty$, or $q_i^a\rightarrow 0$, or $A_i^a\rightarrow \infty$. Consequently,
\begin{align}
\underline{\tau}^a_i=0
\end{align}
%
\subsubsection{Upper Bound of $\tau_i^a$}
Let $\bar {\tau}_i^a$ be the upper bound of $\tau_i^a$. To calculate $\bar {\tau}_i^a$, we need to make the denominator of $\tau_i^a$ in (\ref{Tau_In_variables}) as small as possible. Thus, we set $L_i^a=0$ (or $q_i^a=1$) and $W_{\textrm{min},i}^a=0$, which make the first and fourth terms in the denominator zero. Then, by setting $A_i$ to its lower bound (i.e., $A_i^a=1$), we can minimize the second term in the denominator of $\tau_i^a$. Consequently, at $L_i^a=0,W_{\textrm{min},i}^a=0$, and $A_i^a=1$,
\begin{align}
\tau_i^a=\left[1+\frac{(1+p_i^aN)(2-p_i^a)}{(1-p_i^a)(1-{(p_i^a)}^{m_i+h_i+1})}\right]^{-1}\label{tau_upper_bound-1}
\end{align}
From (\ref{tau_upper_bound-1}), it is clear that the upper bound of $\tau_i^a$ can be achieved when $m_i^a+h_i^a\rightarrow \infty$. Thus, $\bar {\tau}_i^a$ can be written as 
\begin{align}
\bar {\tau}_i^a=\left [1+\frac{(1+p_i^aN)(2-p_i^a)}{1-p_i^a}\right ]^{-1}\label{Tau_Upper_Bound}
\end{align}
\subsection{Optimization Problem}
Taking into account the feasibility region of $\tau_i^a$ derived in Section \ref{SubSection:Feasibility_Region}, here, we can mathematically present the transmission probability optimization problem. The objective is to maximize the overall network throughput, while distributing access airtime among different ISPs according to their reservations. More specifically, the optimization can be formulated as
\begin{subequations}\label{Org-opt-Mod}
\begin{align}
& \max_{\pmb{X,P}} \sum_{i \in \mathcal{S},a \in \mathcal{A}} \frac{x_i^a r_i^a t}{\prod_{i' \in \mathcal{S }}(1+x_{i'}^a)-t'}, ~~~\mathrm{sub}\mathrm{ject}\ \mathrm{to},\label{No-Qos-DC1-Mod} \\
&\sum_{i \in \mathcal{S}_k, a \in \mathcal{A}} \frac{x_i^a \prod_{i' \in \mathcal{S}, i' \neq i}(1+x_{i'}^a)}{\prod_{i' \in \mathcal{S}}(1+x_{i'}^a)-t'} \geq \eta_k, \ \forall k \in \mathcal{K} \label{No-Qos-DC2-Mod} \\
& \frac{x_i^a}{1+x_i^a}\leq \bar{\tau}_i^a,~\forall i\in \mathcal{S},a\in \mathcal{A}\label{No-Qos-DC3-Mod}\\
& \bar{\tau}_i^a\left(1+\frac{(1+p_i^aN)(2-p_i^a)}{1-p_i^a}\right)=1,~\forall i\in \mathcal{S},a\in \mathcal{A}\label{No-Qos-DC4-Mod}\\
&p_i^a=1-\prod_{i^\prime\in \mathcal{S},i^\prime\neq i}{1-\frac{x_{i^\prime}^a}{1+x_{i^\prime}^a}},~\forall i\in \mathcal{S},a\in \mathcal{A}.\label{No-Qos-DC5-Mod}
\end{align}
\end{subequations}
where $\pmb{X}=[x_i^a]$ ($x_i^a \geq 0$) and $\pmb{P}=[p_i^a]$ ($0 \leq p_i^a \leq 1$). Let us recall that $x_i^a=\frac{\tau_i^a}{1-\tau_i^a}$ and $p_i^a=1-\prod_{i^{\prime}\neq i}{(1-\tau_{i^{\prime}}^a)}$. It should be noted that the optimization problem is alternatively formulated with respect to $x_i^a$ instead of $\tau_i^a$, since it will prove useful to solve the problem. 

In the preceding optimization problem, the objective function in (\ref{No-Qos-DC1-Mod}) represents the overall network throughput (i.e., $\sum_{i \in \mathcal{S}, a \in \mathcal{A}}T_i^a$) based on (\ref{Eq:Throughput}). Furthermore, constraints in (\ref{No-Qos-DC2-Mod}) guarantee the minimum airtime reservations for all ISPs (i.e., $\sum_{i \in \mathcal{S}_k, a \in \mathcal{A}} T_{\mathrm{air},i}^a \geq \eta_k $) based on (\ref{Eq:Airtime2}). This set of constraints enable controlling ISPs' share of access airtime regardless of their number of STAs. To guarantee that the provided solution fits into the feasibility region, in constraint (\ref{No-Qos-DC3-Mod})$, \tau_i^a$ is limited to its upper bound. Then, the equality constraint (\ref{No-Qos-DC4-Mod}) establishes the relationship between the upper bound $\bar{\tau}_i^a$ and $p_i^a$ based on (\ref{Tau_Upper_Bound}). Finally, constraint (\ref{No-Qos-DC5-Mod}) establishes the relationship between $p_i^a$ and all $x_{i^\prime}^a,~\forall i^\prime \in \mathcal{S},~ i^\prime\neq i$ according to (\ref{Eq:Pi}).

The formulated problem is non-convex and thus intractable to solve. However, it potentially looks like an extension of Geometric Programming (GP) (defined in Section \ref{Section_Monomial_Approximation}). Thus, by applying successive transformation strategies, we will try to convert the original problem into a series of standard GP problems that can be solved to reach an optimal solution. First, we introduce three auxiliary variables, $y^a=\prod_{i' \in \mathcal{S}}(1+x_{i'}^a)-t',\ \forall a \in \mathcal{A}$, $u_i^a=1-p_i^a, \ \forall i \in \mathcal{S}, \forall a \in \mathcal{A}$, and $t_i^a=1+x_i^a, \ \forall i \in \mathcal{S}, \forall a \in \mathcal{A}$. Then, the optimization problem in (\ref{Org-opt-Mod}) can be transformed into
\begin{subequations}
\begin{align}
& \min_{\pmb{X},\pmb{T},\pmb{U},\pmb{P},\pmb{Y}} \sum_{i \in \mathcal{S},a \in \mathcal{A}} -\frac{x_i^a r_i^a t}{y^a}, ~~~\mathrm{sub}\mathrm{ject}\ \mathrm{to},\label{Optimization:auxiliary-1} \\
&\frac{\prod_{i \in \mathcal{S}}(1+x_i^a)}{t'+y^a}=1, \ \forall a \in \mathcal{A}  \\
&\frac{\eta_k+1}{1+\sum_{i \in \mathcal{S}_k, a \in \mathcal{A}}\frac{x_i^a \prod_{i' \neq i} t_i^a}{y^a}} \leq 1 , \ \forall k \in \mathcal{K} \\
&\frac{x_i^a}{1+x_i^a}\leq\frac{1}{1+\frac{(1+p_i^aN)(2-p_i^a)}{1-p_i^a}},~\forall i\in \mathcal{S},a\in \mathcal{A}\\
& u_i^a\prod_{i^\prime\in\mathcal{S},i^\prime\neq i}{t_{i^\prime}^a}=1,~\forall i\in \mathcal{S},a\in \mathcal{A}\\
&\frac{t_i^a}{1+x_i^a}=1,~\forall i\in \mathcal{S},a\in \mathcal{A}\\
&u_i^a+p_i^a=1,~\forall i\in \mathcal{S},a\in \mathcal{A}\label{Optimization:auxiliary-7}
\end{align}
\end{subequations}
where $\pmb{T}=[t_i^a]$ ($t_i^a \geq 1$), $\pmb{U}=[u_i^a]$ ($0 \leq u_i^a \leq 1$), $\pmb{Y}=[y^a]$ ($y^a > 0$). Nevertheless, the transformed problem is not still in a GP form. One reason is that the objective function (\ref{Optimization:auxiliary-1}) is not a posynomial because of negative multiplicative coefficients. To deal with such problem, first, we equivalently substitute the objective function by $\sum_{i \in \mathcal{S},a \in \mathcal{A}} -x_i^a r_i^a t (y^a)^{-1}+M$ where $M$ is a sufficiently large positive constant. Adding $M$ makes sure that the objective function is always positive. Then, we introduce an additional auxiliary variable $x_0 \geq 0$. By minimizing $x_0$ and guaranteeing constraint C11 in (\ref{Eq:CGP}), we can effectively minimize the objective function in (\ref{Optimization:auxiliary-1}). Consequently, 
\begin{subequations}\label{Eq:CGP}
\begin{align}
& \min_{\pmb{X},\pmb{T},\pmb{U},\pmb{P},\pmb{Y}, x_0} \ x_0, ~~~\mathrm{sub}\mathrm{ject}\ \mathrm{to},  \\
&\mathrm{C}11:\frac{M}{x_0+\sum_{i \in \mathcal{S},a \in \mathcal{A}}\left(\frac{x_i^a r_i^a t}{y^a}\right)}\leq 1 \nonumber \\
&\mathrm{C}12:\frac{\prod_{i \in \mathcal{S}}(1+x_i^a)}{t'+y^a}=1, \ \forall a \in \mathcal{A} \nonumber \\
&\mathrm{C}13:\frac{\eta_k+1}{1+\sum_{i \in \mathcal{S}_k, a \in \mathcal{A}}\frac{x_i^a \prod_{i'\in \mathcal{S},i' \neq i} t_{i^{\prime}}^a}{y^a}} \leq 1 , \ \forall k \in \mathcal{K} \nonumber\\
&\mathrm{C}14:\frac{u_i^ax_i^a+(1+N)x_i^a}{u_i^a+N(u_i^a)^2x_i^a}\leq 1,~\forall i\in \mathcal{S},a\in \mathcal{A}\nonumber\\
&\mathrm{C}15:u_i^a\prod_{i^\prime\in\mathcal{S},i^\prime\neq i}{t_{i^\prime}^a}=1,~\forall i\in \mathcal{S},a\in \mathcal{A} \nonumber\\
&\mathrm{C}16:\frac{t_i^a}{1+x_i^a}=1, \ \forall i\in \mathcal{S},a\in \mathcal{A} \nonumber
\end{align}
\end{subequations}
In the preceding optimization problem, $p_i^a$ is replaced by $1-u_i^a$ based on constraint in (\ref{Optimization:auxiliary-7}). The optimization problem in (\ref{Eq:CGP}) belongs to the class of \textit{complementary GP} problems that allow upper bound constraints on the ratio between two posynomials and equality constraints on the ratio between a monomial and a posynomial \cite{chiang2005geometric,Xu_2014_Signomial}. By approximating the posynomials in the denominator of such constraints, a \textit{complementary GP} can be turned into a standard form of GP. Consequently, the optimal solution can be achieved by iteratively applying monomial approximations and solving a series of GPs. The arithmetic-geometric mean inequality can be used to approximate a posynomial with a monomial. The details of such monomial approximation are provided in Section \ref{Section_Monomial_Approximation}.

Accordingly, we propose an iterative algorithm to reach to an optimal solution of the transmission probability optimization problem. In each iteration, monomial approximations are applied to the denominator of C11, C12, C13, C14, and C16. Then, the resulting GP can be solved for instance by using a standard interior-point algorithm. More specifically, Algorithm \ref{Alg:Optimization} presents different steps need to be performed until convergence.
\begin{algorithm}[t]\caption{\textbf{: GP-based Association Control Algorithm}}\label{Alg:Optimization}
\begin{spacing}{0.85}
\begin{algorithmic}
\STATE {
Initialize $x_i^a$, $t_i^a$, $p_i^a$, $u_i^a$, for all $i\in \mathcal{S}, a\in \mathcal{A}$, $y_a$ for all $a\in\mathcal{A}$, $x_0$};\\
Record the current system state as $\pmb{Z}=(\pmb{X,T,U,P,Y},x_0)$;
\REPEAT
    \STATE {
    Compute the ratio $\alpha$ of each monomial term in the denominator of $\textrm{C11}$, $\textrm{C12}$, $\textrm{C13}$, $\textrm{C14}$, and $\textrm{C16}$ according to (\ref{Defintion_Alpha}), at the current system state $\pmb{Z}$;\\
	Apply monomial approximation to the denominators mentioned above according to (\ref{Monomial_Approximation});\\
	Solve the resulting GP problem using cvx;\\
	Update the current system state $\pmb{Z}=(\pmb{X,T,U,P,Y},x_0)$;}
\UNTIL{all $x_i^a$ converge.}
\end{algorithmic}
Compute the optimal transmission probabilities ${\tau^*}_i^a=\frac{{x^*}_i^a}{1+{x^*}_i^a}$
\end{spacing}
\vspace{-0mm}
\end{algorithm}

\subsection{Asymptotic Complexity Analysis}
In this subsection, we study the asymptotic complexity {\color{black}and scalability} of Algorithm \ref{Alg:Optimization} in terms of the number of STAs (i.e., $N_s$) and the number of APs (i.e., $N_a$).

In Algorithm \ref{Alg:Optimization}, for each iteration, the computational complexity is incurred by applying monomial approximations and solving the resulting GP problem. Suppose $C_\mathrm{MA}$ and $C_\mathrm{GP}$ denote the required computing efforts for monomial approximations and solving GP problem in each iteration.

More specifically, first in each iteration, the denominators of constraints C11, C12, C13, C14, and C16 need to be approximated according to (\ref{Monomial_Approximation}). The required computational complexity for these monomial approximations is proportional to the number of monomial terms in the denominators of constraints C11, C12, C13, C14 and C16. Consequently, the complexity of monomial approximations totals to
\begin{align}\label{Eq:MA}
C_\mathrm{MA}=&\mathcal{O}(N_aN_s)+\mathcal{O}(N_aN_s)+\mathcal{O}(N_aN_s^2)\nonumber\\
&+\mathcal{O}(N_aN_s)+\mathcal{O}(N_aN_s)
=\mathcal{O}(N_aN_s^2)
\end{align}
Subsequently, the resulting GP problem needs to be solved by transforming it to a convex problem. It is reported in \cite{2001_Hershenson_Computer} that the worst-case computational complexity of this approach is $\mathcal{O}(pn^3)$, where $n$ is the number of variables, and $p$ is the total number of terms in all monomials and posynomials in the objective function and constraints. The number of variables of the optimization problem in (\ref{Eq:CGP}) is
\begin{align}\label{GP_variables}
n=4N_aN_s+2N_a+1=\mathcal{O}(N_aN_s)
\end{align}
Furthermore, the total number of terms in all monomials and posynomials can be counted as
\begin{align}\label{GP_term_Counter}
p= & 1+(4N_aN_s+1)+N_a(N_s+2)\nonumber\\
+ & \sum_k(1+N_k N_a(N_s+1)) \nonumber \\
+& 7N_aN_s+N_aN_s^2+2N_aN_s \nonumber\\
= & \mathcal{O}(N_aN_s^2)
\end{align}
where each term in (\ref{GP_term_Counter}) is respectively the number of terms in all the monomials and posynomials in the objective function and constraints C11-C16. Based on (\ref{GP_variables}) and (\ref{GP_term_Counter}), to solve the GP problem, the complexity is
\begin{align}\label{Eq:GP}
C_\mathrm{GP}=\mathcal{O}(pn^3)=\mathcal{O}(N_a^4N_s^5)
\end{align}
Thus, based on (\ref{Eq:MA}) and (\ref{Eq:GP}), \textit{per-iteration asymptotic complexity} of Algorithm \ref{Alg:Optimization} becomes
\begin{align}\label{complexity}
C_{\mathrm{Alg} \ref{Alg:Optimization}\mathrm{-iteration}}=C_\mathrm{MA}+C_\mathrm{GP}
=\mathcal{O}(N_a^4N_s^5)
\end{align}
\begin{figure}[!t]
\centering
\subfloat[Number of iterations vs. number of APs ($N_a$)]{
        \label{subfig:Complexity_a}
        \includegraphics[width=0.48\textwidth]{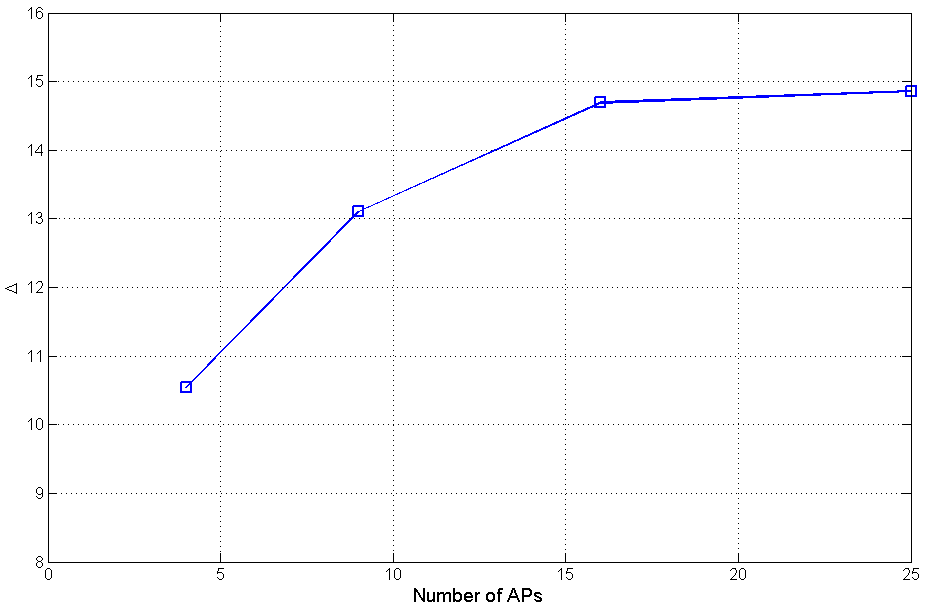} } \\
\subfloat[Number of iterations vs. STA density ($\lambda_\text{mean}$)]{
        \label{subfig:Complexity_b}
        \includegraphics[width=0.48\textwidth]{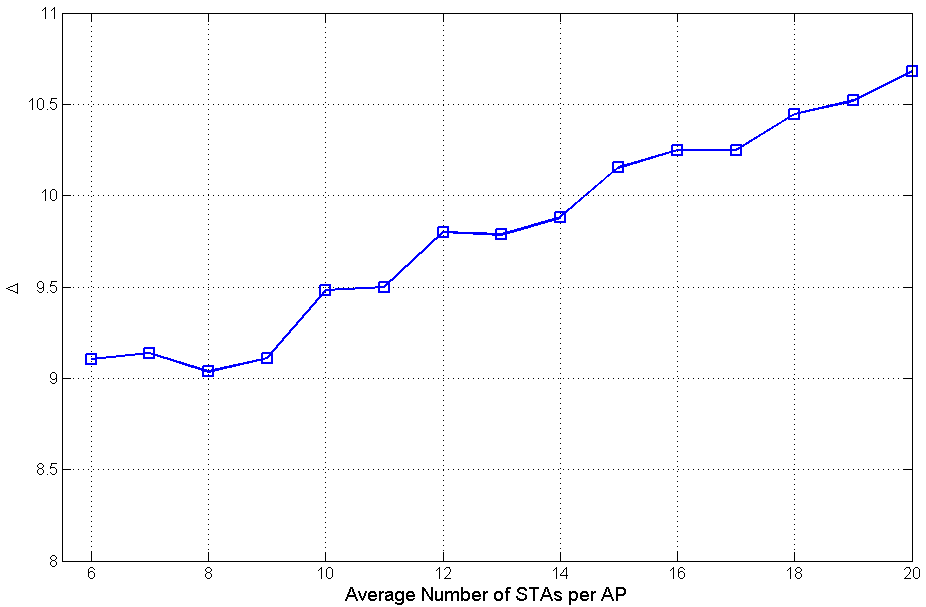} } 
\caption{Number of iterations required for Algorithm \ref{Alg:Optimization} to converge}
\label{Fig:Complexity_Alg:Optimization}
\end{figure}
We now turn to studying the number of iterations (denoted by $\Delta$) required for Algorithm \ref{Alg:Optimization} to converge. Figures \ref{subfig:Complexity_a} and \ref{subfig:Complexity_b} illustrate numerical results on $\Delta$ versus number of APs (i.e., $N_a$) and average number of STAs per AP (i.e., $\lambda_\text{mean}$), respectively. The simulation setup used in these figures is the same as the one presented in Section \ref{Section:Illustrative-Results}. In Figure \ref{subfig:Complexity_a}, $\lambda_\text{mean}$ is set to $\frac{10}{N_a}$, so that the expected total number of STAs in the WLAN stays fixed. Figure \ref{subfig:Complexity_a} shows that $\Delta$ increases with $N_a$ when $N_a$ is small, and then, it becomes \textit{steady} for a larger range of $N_a$. In Figure \ref{subfig:Complexity_b}, $N_a$ is fixed equal to $4$ and $\lambda_\text{mean}$ is varying. Figure \ref{subfig:Complexity_b} shows that $\Delta$ grows \textit{linearly} over a typical range of $\lambda_\text{mean}$. It should be noted that the fluctuations in $\Delta$ are caused by the randomness in the number of STAs per AP, which follows a Poisson distribution with mean of $\lambda_\text{mean}$.

Therefore, based on (\ref{complexity}) and the numerical results in Figures \ref{subfig:Complexity_a} and \ref{subfig:Complexity_b}, it is shown that the overall complexity of the proposed algorithm only grows \textit{polynomially} with the number of STAs and APs, compared to an exponential complexity required by direct search methods.
%
%

\section{Implementation Details}\label{Section:Implementation-and-Numerical-Result}
The optimal transmission probability of STA $i$ at AP $a$ (i.e., ${\tau^*}_i^a$) can be obtained from Algorithm \ref{Alg:Optimization}. But, in the EDCA protocol, the transmission probabilities of STAs are not directly controllable. Instead, what we can control are the MAC-layer parameters (e.g., $W_\textrm{min}$ and $\text{AIFS}$) to achieve the optimal performance. In this section, we first verify the accuracy of the relationship between $\tau_i^a$ and the EDCA parameters provided in (\ref{Tau_In_variables}) and also the validity of the Markov chain model. Then, we develop an algorithm to adjust EDCA parameters aiming to implement ${\tau^*}_i^a$ in the EDCA protocol.

{
\color{black}
We first investigate the achievable accuracy by controlling different parameters through an example. In this example, we consider one BSS with $6$ STAs. The transmission probabilities of STA $2$ to STA $6$ are fixed (e.g., $\tau_2=\tau_3=\tau_4=\tau_5=\tau_6=0.005$), while the transmission probability of STA $1$ (i.e., $\tau_1$) is varied from $0.005$ to $0.1$. 
}

\begin{figure}[!t]
\centering
\includegraphics[width=1\linewidth]{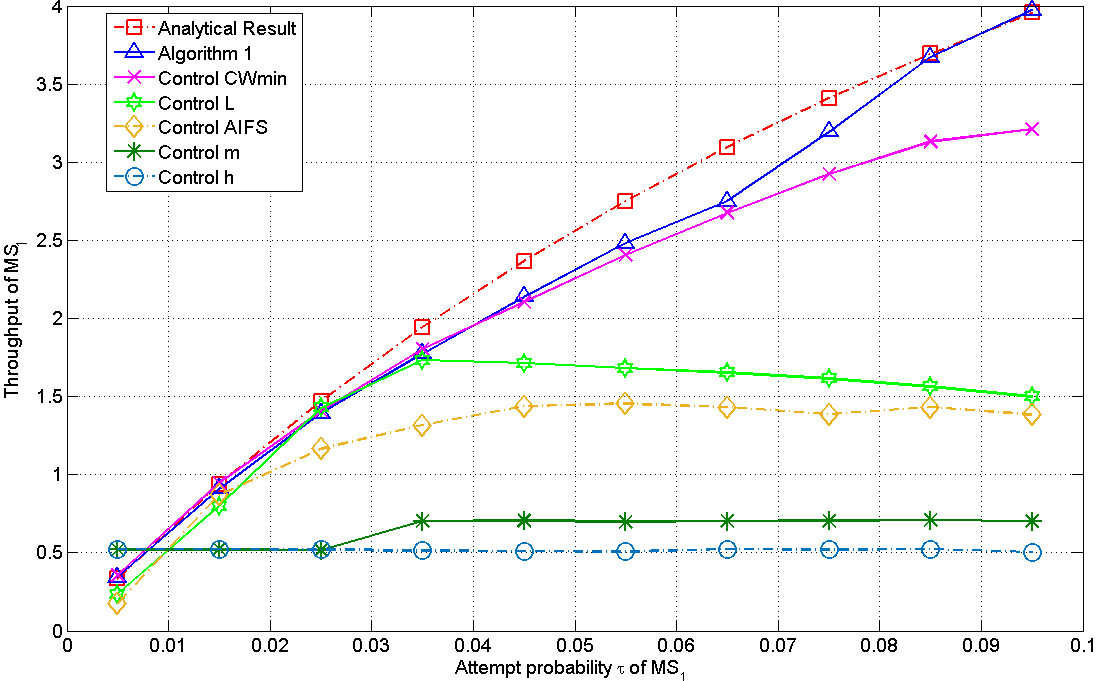} 
\caption{Accuracy of controlling different parameters}
\label{Fig:Accuracy_Different_Par}
\vspace{0 mm}
\end{figure}

Figure \ref{Fig:Accuracy_Different_Par} compares the throughput of STA $1$ analytically derived from (\ref{Eq:Throughput}) and numerically measured using an EDCA simulator. In the numerical results, different EDCA parameters are separately adjusted to tune $\tau_1$ to the desired value. From analytical result in Figure \ref{Fig:Accuracy_Different_Par}, it is clear that the throughput of STA 1 increases with its transmission probability $\tau_1$. Regarding the numerical results, the following conclusions can be drawn.
\begin{algorithm}[t]\caption{\textbf{: MAC Parameter Control Algorithm}}\label{Alg:Parameter_Control}
\begin{spacing}{0.85}
\begin{algorithmic}
\STATE {
\textbf{For each STA-AP pair ($i$, $a$):}\\
$p_i^a=\prod_{i^{\prime} \in \mathcal{S},i^{\prime}\neq i}(1-{\tau^*}_{i^{\prime}}^a)$\\
\textbf{Initialization:}\\
Set $W_{\textrm{min},i}^a=15$, $A_i^a=6$, $q_i^a=0.5$, $L_i^a=100$, $m_i^a=6$, $h_i^a=6$;
}
\STATE{Compute $W_{\textrm{min},i}^a$ from (\ref{Tau_In_variables})}, while other variables are fixed; Round $W_{\textrm{min},i}^a$ to an integer;
\IF{$W_{\textrm{min},i}^a<0$}
\STATE{Set $W_{\text{min},i}^a=0$; }Compute $L_i^a$ from (\ref{Tau_In_variables}), while other variables are fixed;
	   
	   \IF{$L_i^a<0$}
	   \STATE{Set $L_i^a=0$;
	          Compute $A_i^a$ from (\ref{Tau_In_variables}), while other variables are fixed;
	          Round $A_i^a$ to an integer;}
	          \IF{$A_i^a<1$}
	          \STATE{Set $A_i^a=1$;
	          Compute $m_i^a$ from (\ref{Tau_In_variables}), while other variables are fixed;
	          Round $m_i^a$ to an integer;}
	          \IF{$m_i^a<0$}
	          \STATE{Set $m_i^a=0$;
	          Compute $h_i^a$ from (\ref{Tau_In_variables}), while other variables are fixed; Round $h_i^a$ to an integer;}
	          \IF{$h_i^a<0$}
	          \STATE{Set $h_i^a$=0};
	          \ENDIF
	          \ENDIF
	          \ENDIF
	   \ENDIF 
\ENDIF
Start transmission using EDCA protocol with parameters \{$W_{\textrm{min},i}^a$, $A_i^a$, $q_i^a$, $L_i^a$, $m_i^a$, $h_i^a$\}.
\end{algorithmic}
\end{spacing}
\vspace{-0mm}
\end{algorithm}
%

{
\color{black}
From Figure $\ref{Fig:Accuracy_Different_Par}$, ${W_{\textrm{min},i}}$ is the best single parameter to be controlled aiming to adjust $\tau_i$ to a desired value. $A_i$ and $L_i$ provide a good match and thus can be good choices to tune $\tau_i$. Furthermore, $m_i$ and $h_i$ control is not adequately effective, especially for $h_i$. Also, as we can observe, the achievable ranges of $\tau$ by controlling each parameter (i.e., $A_i$, $W_{\text{min},i}$ and $L_i$) are limited. More specifically, when $A_i$ is being adjusted, we can learn from Figure \ref{Fig:Accuracy_Different_Par} that the simulation result matches with the analytical result when $\tau_1$ is in the range $[0.005,0.15)$. When CWmin is being controlled, the situation is similar, but for a broader range of $\tau_i$ ($[0.005,0.065)$). Figure \ref{Fig:Accuracy_Different_Par} also suggests that controlling $L$ is effective when $\tau_1$ is in the range $[0.005,0.035)$. Thus, in summary, to achieve a boarder feasible range of $\tau$, we need to jointly adjust these parameters. 
}

Here, we aim to develop a MAC parameter control algorithm to achieve the optimal transmission probability obtained from the optimization problem in Section \ref{Section-Optimization-Problem}. In Section \ref{SubSection:Feasibility_Region}, the feasibility region of transmission probability is calculated assuming that all the $5$ MAC parameters (i.e., ${W_{\textrm{min}}}$, $A$, $q$ or $L$, $m$, $h$) can be freely varied in their feasible ranges. Thus, a control algorithm is required which can be able to adjust all the $5$ parameters simultaneously. Thus, an algorithm is proposed taking into account the achievable accuracy by controlling different parameters. Algorithm \ref{Alg:Parameter_Control} shows the complete control procedure. 

{
\color{black}
The achievable throughput for STA $1$ in Algorithm \ref{Alg:Parameter_Control} is also plotted in Figure \ref{Fig:Accuracy_Different_Par}. As expected, we can learn that Algorithm \ref{Alg:Parameter_Control} can approach the desired throughput closely for the total range of $\tau_1$ (i.e., the upper bound $\tau_1$ computed from (\ref{Tau_Upper_Bound})). Thus, Algorithm \ref{Alg:Parameter_Control} can improve the control accuracy compared with controlling the parameters separately.
}

\section{Numerical Results}\label{Section:Illustrative-Results}
In this section, we present numerical results to evaluate the performance of the proposed STA-AP association and airtime control algorithm, and also the MAC parameter control algorithm. More specifically, the performance of our GP-based association scheme is compared with the Max-SNR scheme in terms of throughput and fairness. We implemented a simulator for EDCA (including access probability $q$ and inter-frame space $L$ described in Section \ref{Section_Markov_Chain}) in Matlab to measure the achieved throughput numerically. Furthermore, we used cvx to solve the GP problems in the association algorithm. 

We consider a network in which 4 APs are deployed in a $10\times 10 ~m^2$ area. More specifically, the APs are placed at the centers of four different $5\times 5 ~m^2$ grids to provide seamless coverage. To eliminate interference between the transmission of different APs, four non-overlapping $20~\text{MHz}$ channels are assigned to four APs. The STAs are distributed in the entire area according to the two-dimensional Poisson point process (PPP).
%
%
\begin{table}[!t] 
\centering
\begin{tabular}{|c|c|c|c|}
  \hline
  Modulation & FEC Rate & Data Rate (Mbps) & SNR (dB) \\
  \hline
  BPSK & 1/2 &6 &[5,8) \\
  \hline
  BPSK & 3/4 &9 &[8,10)\\
  \hline
  QPSK & 1/2 &12&[10,13) \\
  \hline
  QPSK & 3/4& 18 &[13,16)\\
  \hline
  16QAM & 1/2 & 24&[16,19)\\
  \hline
  16QAM &3/4&36&[19,22)\\
  \hline
  64QAM &2/3&48 &[22,25)\\
  \hline
  64QAM &3/4&54&[25,$\infty$)\\
  \hline  
\end{tabular}
\caption{IEEE 802.11a adaptive modulation and coding scheme and the SNR ranges used in the numerical results}
\label{Table:802.11a_PHY}
\vspace{-2mm}
\end{table}

The wireless channel model includes path loss and small-scale fading. Generally, the channel gain can be expressed as $h=Ah^\prime d^{-\alpha/2}$, where $d$ is the distance between a STA and an AP, $\alpha\geq 2$ is the path loss exponent, $A$ is a constant dependent on the frequency and transmitter/receiver antenna gain, and $h^\prime$ represents the small scale fading component. In the numerical results, we set $\alpha=3$ and $A=1$. Furthermore, $h^\prime$ is randomly generated according to the Rayleigh distribution assuming $\mathbb{E}\{|h^\prime|^2\}=1$. The received SNR at STA $i$ is equal to $\frac{Pg_i^a}{\sigma^2}$ where $P$ is the transmission power, $g_i^a=|h_i^a|^2$ is the channel power gain from STA $i$ to AP $a$, and $\sigma^2$ is the power of noise. In all the numerical results, $P/\sigma^2$ is assumed fixed and set to $10\text{dB}$.

To determine the transmission rate of each STA-AP pair, the 802.11a physical layer model is used. More specifically, to guarantee a maximum packet error rate, adaptive modulation and coding is used based on the received SNR. Table \ref{Table:802.11a_PHY} shows the achievable transmission rates and adaptive modulation and coding schemes standardized in IEEE 802.11a and the SNR range for each scheme used in the simulations.

The MAC layer parameters used in our simulations are summarized in Table \ref{Table:802.11e_MAC}. Moreover, the target airtime share for each ISP $k$ (i.e., $\eta_k$) is set equal to the number of APs divided by the number of ISPs. In other words, we assume that the ISPs have the same minimum airtime reservation and share the total airtime in a fair manner. In the Markov chain, the average frozen time $N$ is approximated by $T_{\textrm{TXOP}}/\delta$.
\begin{table}[!t] 
\centering
\begin{tabular}{|c|c|}
  \hline
  time-slot $\delta$ & 9 $\mu s$ \\
  \hline
  Propagation Delay $\gamma$ & 1 $\mu s$ \\
  \hline
  $T_{\textrm{TXOP}}$ & 1 $ms$ \\
  \hline
  SIFS & 10 $\mu s$ \\
  \hline
  ACK & 40 $\mu s$ \\
  \hline
\end{tabular}
\caption{IEEE 802.11e MAC parameters used in the numerical results}
\label{Table:802.11e_MAC}
\vspace{-2mm}
\end{table}
%
%
\begin{figure}[!t]
\centering
\includegraphics[width=0.85\linewidth]{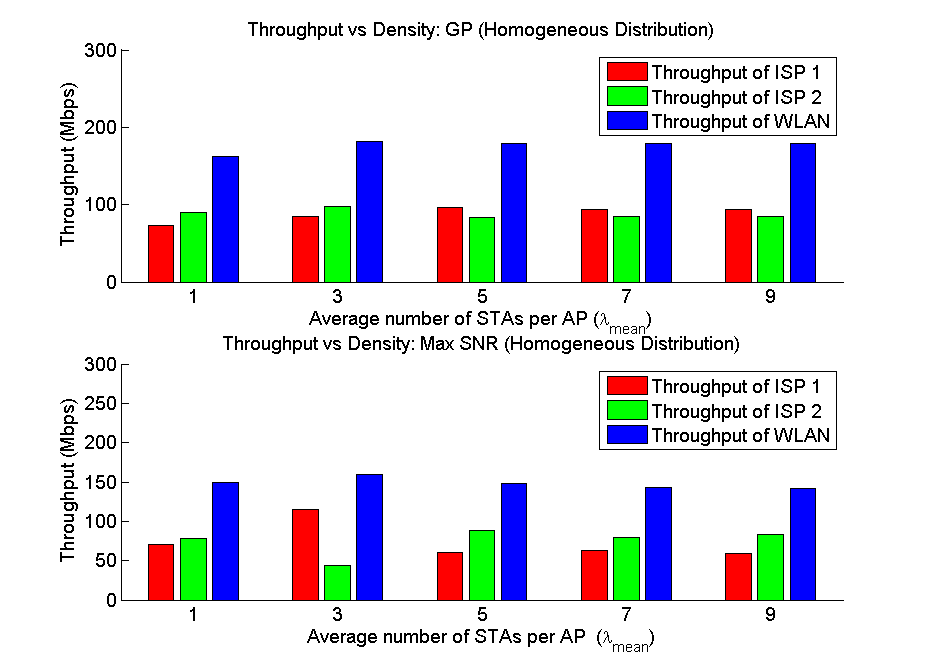}
\caption{Throughput vs. STA density, homogeneous distribution}
\label{Fig:Throuput_vs_Density_Homogeneous}
\vspace{-3mm}
\end{figure}
\begin{figure}[!t]
\centering
\includegraphics[width=0.85\linewidth]{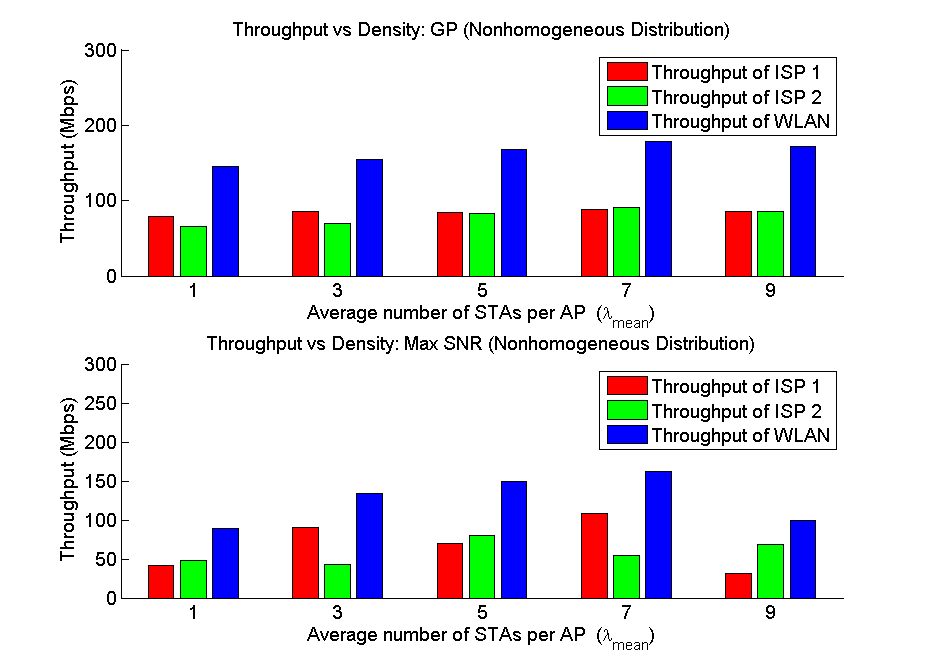}
\caption{Throughput vs. STA density, non-homogeneous distribution}
\label{Fig:Throughput-vs-Density-Nonhomogeneous}
\vspace{-5mm}
\end{figure}
\begin{figure}[!t]
\centering
\includegraphics[width=0.85\linewidth]{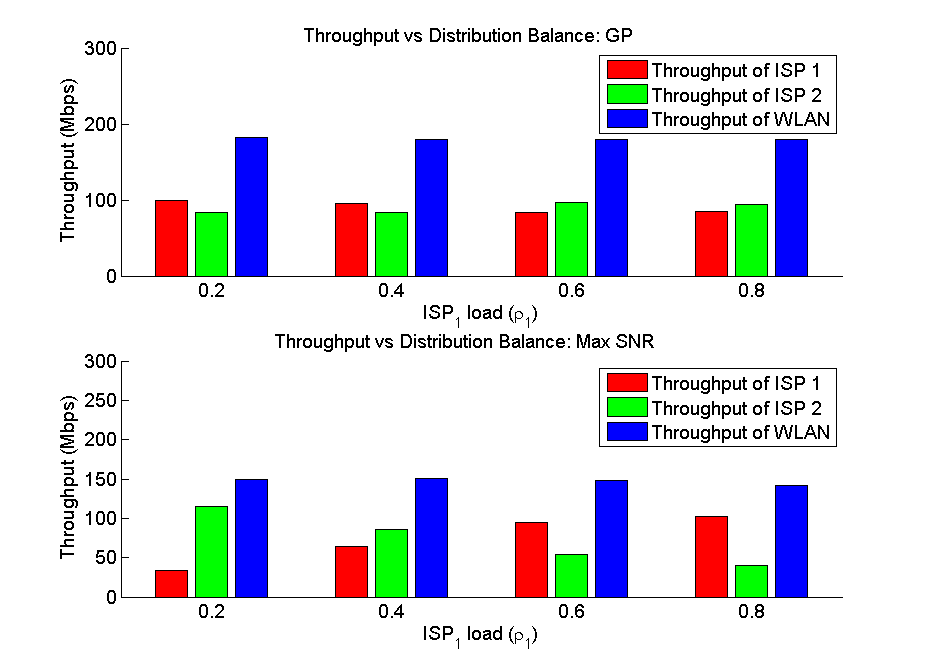}
\caption{Throughput vs. ISP 1 load}
\label{Fig:Throughput-vs-Balance-Bar}
\vspace{-5mm}
\end{figure}
%
\subsection{Effects of STA Distribution}
Here, we investigate the impact of STA distribution on the fairness and throughput achieved by the two association algorithms. More specifically, we set up two examples considering homogeneous and non-homogeneous STA distributions. {\color{black}In both examples, the STAs are randomly associated to the two ISPs with equal probabilities.}

\subsubsection{Homogeneous Distribution}
In this example, the STAs are distributed in the square space according to a homogeneous two-dimensional PPP. Accordingly, the number of STAs in each grid (where an AP centered at) follows a Poisson distribution with mean $\lambda_\text{mean}$, which represents the average number of STAs per AP. Figure \ref{Fig:Throuput_vs_Density_Homogeneous} shows that GP-based association scheme improves both fairness and total throughput compared with Max-SNR scheme.

\subsubsection{Non-homogeneous Distribution} 
Let consider that STAs are distributed according to a non-homogeneous PPP. More specifically, the number of STAs in a grid (where AP $a$ centered at) follows a Poisson distribution with mean $\lambda_a$. Here, $\lambda_a$ is randomly generated between $0$ and $\lambda_\mathrm{mean}$. With such STA distribution, Figure \ref{Fig:Throughput-vs-Density-Nonhomogeneous} shows the achieved throughput of two ISPs versus $\lambda_\text{mean}$. As expected, the performance gap is significantly larger with non-homogeneous distribution compared to the homogeneous case. Max-SNR can hardly guarantee the fairness between different ISPs. While for the same distribution, GP-based STA-AP association can manage to keep the balance between two ISPs. Furthermore, it can offer improvement over Max-SNR in terms of the total throughput due to the load-balancing among APs.

\subsection{Effects of STA Density and ISP Load} 
Let define $\rho_1$ (also referred to as ISP 1 load) as the ratio of number of STAs serving by ISP 1 to the total number of STAs in the network. Here, the performance of the two association approaches are compared under different STA density and ISP load.

Assuming a homogeneous STA distribution with $\lambda_{\text{mean}}=3$, Figure \ref{Fig:Throughput-vs-Balance-Bar} demonstrates the achieved throughput of two ISPs versus different values of $\rho_1$ for both GP-based and Max-SNR association schemes. By Max-SNR association, it is shown that throughput of ISP $1$ grows linearly with $\rho_1$, while the achieved throughput of ISP 2 is decreasing. But, GP-based association can fairly distribute the airtime between ISPs regardless of their ISP loads and thus keep a perfect balance between the achieved throughput of the two ISPs.

\begin{figure}[!t]
\centering
\includegraphics[width=0.95\linewidth]{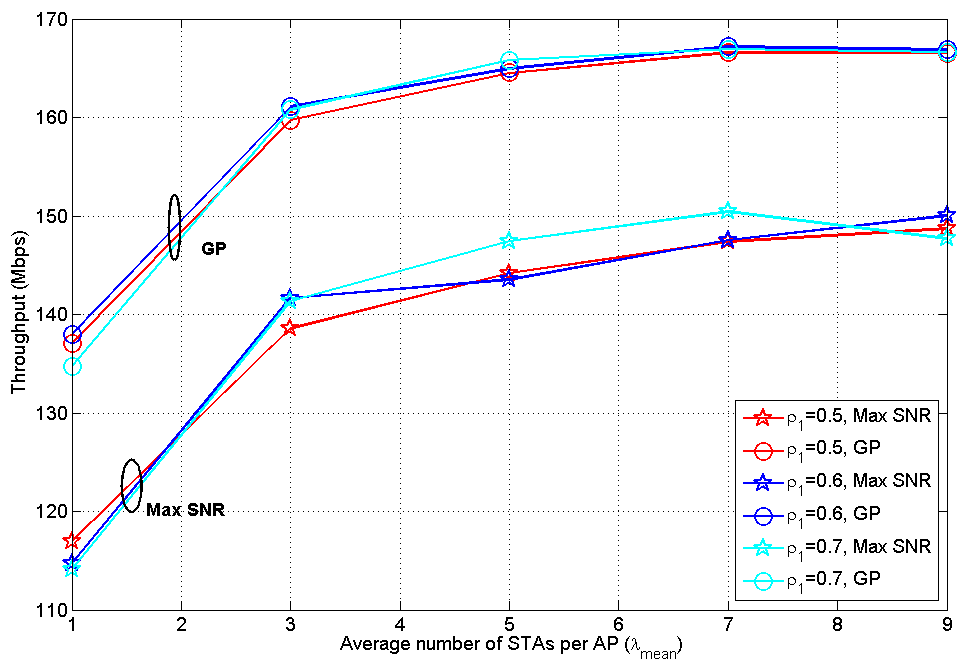}
\caption{Total throughput vs. STA density for different $\rho_1$}
\label{Fig:Throughput_vs_Density_Balance}
\vspace{-3mm}
\end{figure}
\begin{figure}[!t]
\centering
\includegraphics[width=0.95\linewidth]{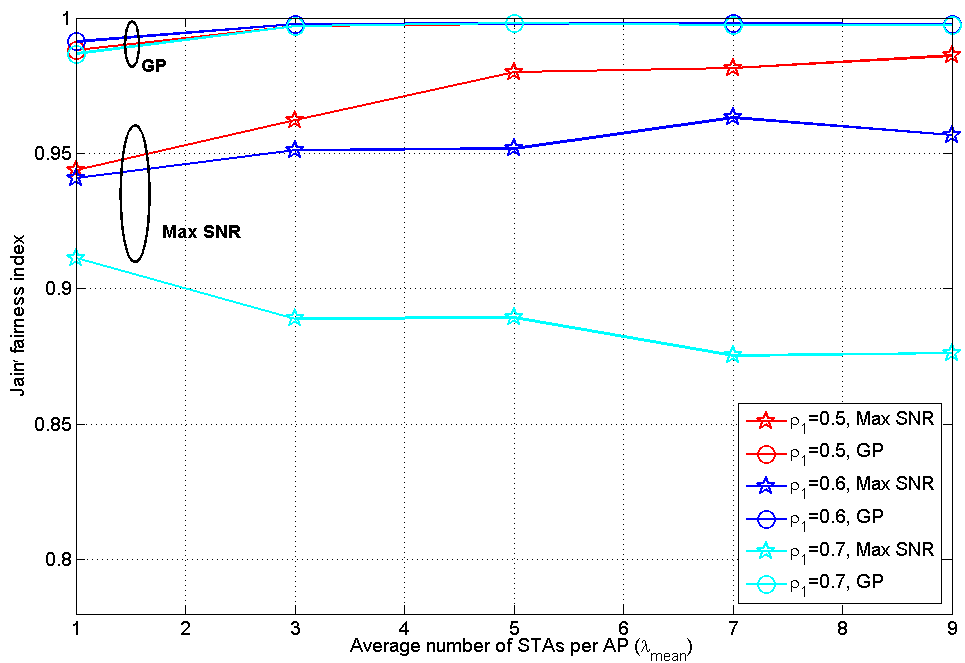}
\caption{Fairness vs. STA density for different $\rho_1$}
\label{Fig:Fairness_vs_Density_Balance}
\vspace{-6mm}
\end{figure}
Figure \ref{Fig:Throughput_vs_Density_Balance} shows the total throughput achieved by the two association algorithms versus $\lambda_{\text{mean}}$ for a homogeneous STA distribution. For a fixed $\rho_1$, the total throughput by both algorithms increases with the STA density (i.e., $\lambda_{\text{mean}}$). But, the throughput increase rate is decreasing with $\lambda_{\text{mean}}$. This is because the wireless channel is underutilized when the STA density is low. Thus, the increase in the STA density will improve the total throughput. But, when the STA density is large, increasing the STA density further will result in a higher collision probability, and hence, slow down the total throughput improvement. For any fixed $\rho_1$, it is shown that GP-based association significantly improves the total throughput compared with the Max-SNR association. 

Figure \ref{Fig:Fairness_vs_Density_Balance} measures the fairness by employing the Jain's fairness index in (\ref{Eq:fairness}). 
\begin{align}\label{Eq:fairness}
F=\frac{(\sum_{k\in \mathcal{K}}T_k)^2}{|\mathcal{K}|\sum_{k\in\mathcal{K}}T_k^2}
\end{align}
where $T_k=\sum_{i\in S_k,a\in A}T_i^a$ is the achieved throughput for all the STAs of ISP $k$. From Figure \ref{Fig:Fairness_vs_Density_Balance}, it is clear that the proposed GP-based association approach can always guarantee perfect fairness between the ISPs regardless of the STA density or $\rho_1$. The achieved fairness level by Max-SNR association is always worse than GP-based, especially when the STA load is highly unbalanced between ISPs (i.e., $\rho_1$ is not close to $0.5$).
\vspace{-1mm}
\section{Conclusion}\label{Section:Conclusion}
This paper considers the STA-AP association and airtime control in virtualized 802.11 networks aiming to provide fairness guarantees among ISPs despite the number of STAs per ISP. First, a three-dimensional Markov chain is developed to model a generalized 802.11e EDCA protocol. This model establishes the relationship between the transmission probability of each STA and the detailed parameters in the MAC protocol. Based on this relationship, the feasibility region of the transmission probabilities are derived. Subsequently, an optimization problem is formulated which can maximize the network throughput, while guaranteeing the fairness between different ISPs. The implementation of the optimal transmission probabilities obtained by successive geometric programming is discussed by controlling the MAC parameters. Extensive numerical results confirm that the Markov chain can accurately describe the MAC protocol. Furthermore, it is verified that the proposed association algorithm can improve the throughput and provide fairness guarantees in virtualized 802.11 WLANs.
\section{Appendix}
\subsection{Markov Chain Model}\label{Section_Markov_Chain}
{
\color{black}
Here, we detail the Markov chain model for EDCA introduced in Section \ref{SubSection:Markov_Chain}. Firstly, the transition rules among the states of the Markov chain are given according to the EDCA protocol. Then, the transition probability between any two adjacent states are summarized. Finally, by solving the stationary point of the Markov chain, the relationship between transmission probability and the detailed EDCA parameters are derived.	
}

\begin{figure*}[!t]
	\centering
	\includegraphics[width=0.9\linewidth]{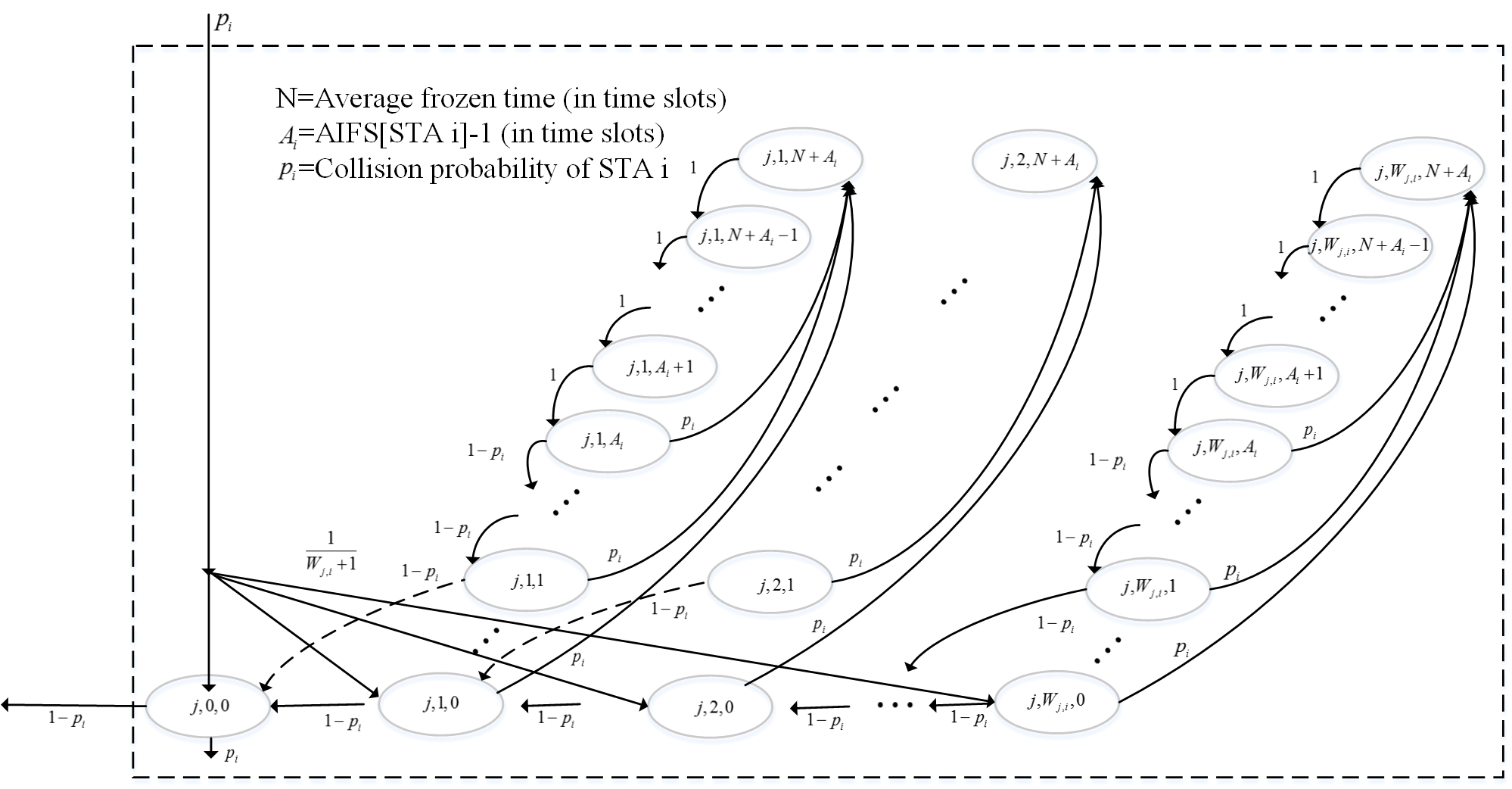}
	\caption{Detailed transition for one backoff stage}
	\label{Fig:Markov_Model_Zoom_In}
	\vspace{-5mm}
\end{figure*}
\subsubsection{Transition rules}\label{Subsection:Markov_Chain}
First, we describe the transitions among the states in the proposed Markov chain according to the EDCA protocol. As shown in Figure \ref{Fig:Markov-Chain}, after a successful transmission or a packet drop, the STA will flip a biased coin with successful probability $q_i$. If the coin appears to be tail (with probability $1-q_i$), the STA will go to the state $\{-2,0,L_i-1\}$, and try again after $L_i$ time-slots. Otherwise, if the coin is head (with probability $q_i$), the STA will move to the state $\{-1,0,A_i\}$, where $A_i=\textrm{AIFS}[\textrm{STA}~i]-1$. If the channel stays idle in the following $A_i+1$ time-slots, the STA will move on to the backoff/retransmission stage $0$ and uniformly pick a random backoff time from $[0,W_{0,i}]$. 

Otherwise, if the STA senses the channel busy in a state $\{-1,0,d\}$ where $0\leq d\leq A_i$, it will freeze and move to the state $\{-1,0,N+A_i\}$, where $N$ is the average frozen time (The frozen time is equal to either $T_s$ or $T_c$ and thus $N$ can be approximated to $T$ as discussed before). Afterwards, it will wait until channel becomes idle again, by counting down $v(t)$ from $N+A_i$ to $A_i$. When the STA gets to the state $\{-1,0,A_i\}$, it has to keep repeating the explained process until it enters the backoff process. We assume that the probability that the channel is observed busy by STA $i$ is a constant $p_i$ in a general time-slot. Given the transmission probabilities of all STAs (i.e., $\tau_i$ ) in the BSS, we have
\begin{align} \label{Eq:Pi}
	p_i=1-\prod_{i^{\prime}\neq i}{(1-\tau_{i^{\prime}})}.
\end{align}
For the backoff process, we use Figure \ref{Fig:Markov_Model_Zoom_In} to explain the possible transitions at the backoff/retransmission stage $j$. When STA $i$ reaches the state $\{j-1,0,0\}$, it will start a transmission. Then, if other STAs happen to access the channel at the same time, STA $i$ will experience a collision. The probability that STA $i$ encounters a collision is equal to $p_i$. In a case of collision, the STA will enter the next backoff/retransmission stage (i.e., $j$), pick a random backoff counter $b$ between $0$ and $W_{j,i}$, and go to the state $\{j,b,0\}$. When the STA reaches the state $\{j,b,0\}$, it will count down to $\{j,b-1,0\}$ if the channel is idle. Otherwise, if the channel is sensed busy (i.e., other STAs are transmitting in the channel), STA $i$ will freeze its backoff counter and move to the state $\{j,b,N+A_i\}$. In the frozen period, the STA will count down $v(t)$ by $1$ each time-slot until it gets to the state $\{j,b,A_i\}$. Then, if the channel is still idle (with probability ($1-p_i$)), the STA will continue counting down to $\{j,b,1\}$ and then go to state $\{j,b-1,0\}$. Otherwise, if the channel becomes busy when STA $i$ is in the state $\{j,b,d\}$ where $1\leq d\leq A_i$, it has to freeze its backoff counter again and go back to the state $\{j,b,N+A_i\}$.

Finally, for states $\{j,0,0\},~0\leq j\leq m_i+h_i$, the STA will initiate a transmission. If no other STA transmits at the same time (with probability ($1-p_i$)), the STA will have a successful transmission of duration $T_s$. Otherwise, if any other STA simultaneously starts a transmission (with probability $p_i$), STA $i$ will experience a collision of duration $T_c$. After the collision, if $0\leq j\leq m_i+h_i-1$, the STA will go to the next backoff/retransmission stage. But, if $j=m_i+h_i$ (which means that the maximum backoff/retransmission limit is reached), the STA has to drop this packet. Then, for a new packet transmission, it has to flip a biased coin and start over from the state $\{-2,0,L_i-1\}$ or $\{-1,0,A_i\}$ according to the result of the coin toss. The mathematical representations of the transition probabilities are summarized in Section \ref{Transition_Probabilities}.  

\subsubsection{Transition Probabilities of the Markov Chain}\label{Transition_Probabilities}
Given the transition rule between the states described in above, we have the following transition probabilities.\\
\textit{1)} For states $\{-2,0,d\},~0\leq d\leq L_i-1$, i.e., after  getting a tail in a coin toss,
\begin{align}
\begin{cases}
P\{(-2,0,d-1)|(-2,0,d)\}=1,~1\leq d\leq L_i-1\nonumber\\
P\{(-2,0,L_i-1)|(-2,0,0)\}=1-q_i\nonumber\\
P\{(-1,0,A_i)|(-2,0,0)\}=q_i\nonumber
\end{cases}
\end{align}
\textit{2)} For states $\{-1,0,d\},~0\leq d\leq N+A_i$, i.e., before entering the backoff process,
\begin{align}
\begin{cases}
P\{(-1,0,d-1)|(-1,0,d)\}=1,~A_i+1\leq d\leq N+A_i\nonumber\\
P\{(-1,0,d-1)|(-1,0,d)\}=1-p_i,~1\leq d\leq A_i\nonumber\\
P\{(-1,0,N+A_i)|(-1,0,d)\}=p_i,~0\leq d\leq A_i\nonumber\\
P\{(0,b,0)|(-1,0,0)\}=\frac{1-p_i}{W_{0,i}+1},~0\leq b\leq W_{0,i}\nonumber
\end{cases}
\end{align}
\textit{3)} For states $\{j,b,d\},~0\leq j\leq m_i+h_i$, i.e., in the backoff process,
\begin{enumerate}
\item[a)] If $b=0,d=0$, i.e., before a successful transmission or a collision,
\begin{align}
\begin{cases}
P\{(-1,0,A_i)|(j,0,0)\}=(1-p_i)q_i,\nonumber \\ ~~~~~~0\leq j\leq m_i+h_i-1~~~~~~~\nonumber\\
P\{(-2,0,L_i-1)|(j,0,0)\}=(1-p_i)(1-q_i),\nonumber \\ ~~~~~~0\leq j\leq m_i+h_i-1\nonumber\\
P\{(j+1,b,0)|(j,0,0)\}=p_i\frac{1}{W_{j+1,i}+1},\nonumber \\
 ~~~~~~0\leq j\leq m_i+h_i-1,~0\leq b\leq W_{j+1,i}\nonumber\\
P\{(-2,0,L_i-1)|(m_i+h_i,0,0)\}=1-q_i\nonumber\\
P\{(-1,0,A_i)|(m_i+h_i,0,0)\}=q_i\nonumber
\end{cases}
\end{align}
\item[b)] If $d=0,1\leq b\leq W_{j,i}$, i.e, in the count down process,
\begin{align}
\begin{cases}
P\{(j,b-1,0)|(j,b,0)\}=1-p_i,~1\leq b\leq W_{j,i}\nonumber\\
P\{(j,b,N+A_i)|(j,b,0)\}=p_i,~1\leq b\leq W_{j,i}\nonumber
\end{cases}
\end{align}
\end{enumerate}

\textit{4)} If $1\leq d\leq N+A_i,~1\leq b\leq W_{j,i}$, i.e., during a frozen period or the AIFS after it,
\begin{align}
\begin{cases}
P\{(j,b,d-1)|(j,b,d)\}=1,~A_i+1\leq d\leq N+A_i\nonumber\\
P\{(j,b,N+A_i)|(j,b,d)\}=p_i,~1\leq d\leq A_i\nonumber\\
P\{(j,b,d-1)|(j,b,d)\}=1-p_i,~2\leq d\leq A_i\nonumber\\
P\{(j,b-1,0)|(j,b,1)\}=1-p_i\nonumber
\end{cases}
\end{align}
	\begin{figure*}[t] 
		\begin{minipage}[t]{\linewidth}
			\begin{@twocolumnfalse}
				\begin{align}
				1&=\sum\limits_{d=0}^{L_i-1}{b_{-2,0,d}}+\sum\limits_{d=0}^{N+A_i}{b_{-1,0,d}}+\sum\limits_{j=0}^{m_i+h_i}{b_{0,0,0}}+\sum\limits_{j=0}^{m_i+h_i}{\sum\limits_{b=1}^{W_{j,i}}{\sum\limits_{d=0}^{N+A_i}{b_{j,b,d}}}}\nonumber\\ 
				&=\left[L_i\frac{1-q_i}{q_i}+\frac{1+p_iN}{p_i}\frac{1-(1-p_i)^{A_i+1}}{(1-p_i)^{A_i+1}}+\frac{1-p_i^{m_i+h_i+1}}{1-p_i}+\frac{1+Np_i}{2(1-p_i)^{A_i}}\sum\limits_{j=0}^{m_i+h_i}{W_{j,i}p_i^j}\right]b_{0,0,0}\label{Normalization_Condition}
				\end{align}
				\vspace{-5mm}
				\begin{center}
					\line(1,0){515}
				\end{center}
			\end{@twocolumnfalse}
		\end{minipage}
	\end{figure*}
\subsubsection{Stationary Condition of the Markov Chain}\label{Stationary_Conditions}
Based on the transition rule above and the structure of the Markov chain, as in \cite{Kong_Makov_802.11e} and \cite{Bianchi2000}, we have,
\begin{align}
&b_{j,0,0}=p_i^jb_{0,0,0},~0\leq j\leq m_i+h_i\nonumber\\
&b_{j,b,0}=\frac{W_{j,i}+1-b}{W_{j,i}+1}b_{j,0,0},~0\leq j\leq m_i+h_i,1\leq b\leq W_{j,i}\nonumber
\end{align}
By exploiting the transition structure within a frozen process and the post AIFS (i.e., the states $\{j,b,d\},~0\leq d\leq N+A_i$), we have,
\begin{align}
&b_{j,b,d}=\frac{p_i}{(1-p_i)^d}b_{j,b,0},0\leq j\leq m_i+h_i-1,1\leq d\leq A_i-1\nonumber\\
&b_{j,b,d}=\frac{p_i}{(1-p_i)^{A_i}}b_{j,b,0},0\leq j\leq m_i+h_i,A_i\leq d\leq N+A_i\nonumber
\end{align}
For states $\{-2,0,d\}$, by looking at the incoming and outgoing probability, we can have
\begin{align}
b_{-2,0,d}=\frac{1-q_i}{q_i}b_{0,0,0},~0\leq d\leq L_i-1\nonumber
\end{align}
For states $\{-1,0,d\}$, by analyzing the structure of transition between the states $\{-1,0,d\},0\leq d\leq N+A_i$ and the state $\{0,0,0\}$, we can have
\begin{align}
&b_{-1,0,d}=\frac{1}{(1-p_i)^{d+1}}b_{0,0,0},~0\leq d\leq A_i\nonumber\\
&b_{-1,0,d}=\frac{1-(1-p_i)^{A_i+1}}{(1-p_i)^{A_i+1}}b_{0,0,0},~A_i+1\leq d\leq N+A_i\nonumber
\end{align}
{\color{black}
\noindent
To this point, all the stationary probabilities in the Markov chain are represented in terms of $b_{0,0,0}$. Then, $b_{0,0,0}$ can be derived from the normalization condition in ({\ref{Normalization_Condition}). Consequently, having all stationary probabilities $b_{j,b,d}$, the transmission probability of STA $i$ can be calculated as in (\ref{Tau_In_variables}). To derive (\ref{Tau_In_variables}), the contention window $W_{j,i}$ is set equal to
	\begin{align}
	W_{j,i}=\begin{cases}
	W_{\textrm{min},i}2^j,~~0\leq j\leq m_i \nonumber\\
	W_{\textrm{min},i}2^{m_i},~~m_i+1\leq j\leq m_i+h_i \nonumber
	\end{cases}
	\end{align}
	according to the exponential backoff rules in 802.11e. From the definition of transmission probability, $\tau_i$ can be calculated by summing all the stationary probabilities of the states in which the STA will initiate a transmission (i.e., $b_{j,0,0},~\forall j\in\{0,1,...,m_i+h_i\}$). The expression in (\ref{Tau_In_variables}) establishes the relationship between $\tau_i$ and all the detailed parameters for STA $i$ in the EDCA protocol, which will be useful to characterize the feasibility region of $\tau_i^a$ in the optimization and design an algorithm to control the transmission probabilities of the STAs.
}}
\subsection{Geometric Programming and Monomial Approximation}\label{Section_Monomial_Approximation}
An optimization problem is called geometric programming if it follows the following form,
\begin{subequations}
\begin{align}
& \min_{\pmb x} f_0(\pmb x), ~~~\mathrm{sub}\mathrm{ject}\ \mathrm{to}\nonumber\\
& f_i(\pmb x)\leq 1, \ i=1,\dots,n_1\nonumber  \\
& g_i(\pmb x)= 1, \ i=1,\dots,n_2   \nonumber
\end{align}
\end{subequations}
where $f_0,\dots,f_{n_1}$ are posynomials and $g_1,\dots,g_{n_2}$ are monomials. In the context of geometric programming, a monomial function $f$ of $\pmb x=(x_1,x_2,\dots,x_n)$ is defined as,
\begin{align}
f(\pmb x)=cx_1^{a_1}x_2^{a_2}\dots x_n^{a_n}\nonumber
\end{align}
where $c>0$ and $a_i\in \mathcal{R}$.
Furthermore, a posynomial is defined as the summation of multiple monomials, i.e.,
\begin{align}
g(\pmb x)=\sum\nolimits_{k=1}^Kf_{k}(\pmb x)\nonumber 
\end{align}
The basic idea of monomial approximation is as follows: consider a posynomial function $g(\pmb x)=\sum_k{f_k(\pmb x)}$ with $f_k(\pmb x)$ being the monomial terms. By the arithmetic-geometric mean inequality, we have
\begin{align}\label{Monomial_Approximation}
g(\pmb x)\geq \hat g(\pmb x)=\prod\limits_{k}\left(\frac{f_k(\pmb x)}{\alpha_k(\pmb x_0)}\right)^{\alpha_k(\pmb x_0)}
\end{align}
where the parameters $\alpha_k(\pmb x_0)$ can be obtained by computing
\begin{align}\label{Defintion_Alpha}
\alpha_{k}(\pmb x_0)=\frac{f_k(\pmb x_0)}{g(\pmb x_0)},~\forall k 
\end{align}
where $\pmb x_0>0$ is a fixed point (e.g., the optimal solution from the last round of optimization). It is proved that $\hat g(\pmb x)$ is the best local monomial approximation of $g(\pmb x)$ near $\pmb x_0$ \cite{2004_Boyed_ConvexOptimization}.


\end{document}